\documentclass[%
 aip,
 amsmath,amssymb,
 reprint,%
]{revtex4-1}

\usepackage{graphicx}
\usepackage{dcolumn}
\usepackage{bm}

\usepackage[utf8]{inputenc}
\usepackage[T1]{fontenc}
\usepackage{mathptmx}
\usepackage{etoolbox}

\usepackage[colorlinks=true, allcolors=blue]{hyperref}
\usepackage{multirow}
\usepackage{float}
\usepackage{mathtools} 
\usepackage[dvipsnames]{xcolor}
\usepackage{tikz}
\newcommand{\tikzcircle}[2][red,fill=red]{\tikz[baseline=-0.5ex]\draw[#1,radius=#2] (0,0) circle ;}

\newcommand{\hlf}{\frac{1}{2}}
\newcommand{\lt}{\left}
\newcommand{\rt}{\right}

\makeatletter
\def\@email#1#2{%
 \endgroup
 \patchcmd{\titleblock@produce}
  {\frontmatter@RRAPformat}
  {\frontmatter@RRAPformat{\produce@RRAP{*#1\href{mailto:#2}{#2}}}\frontmatter@RRAPformat}
  {}{}
}%
\makeatother
\begin{document}

\preprint{AIP/123-QED}

\title[Island myriads in periodic potentials]{Island myriads in periodic potentials}

\author{Matheus J. Lazarotto}
 \affiliation{Aix-Marseille Universit\'{e}, CNRS, UMR 7345 PIIM, F-13397, Marseille cedex 13, France}
 \affiliation{Instituto de F\'{i}sica, Universidade de S\~{a}o Paulo, Rua do Mat\~{a}o 1371, S\~{a}o Paulo 05508-090, Brazil}
 
\author{Iber\^{e} L. Caldas}
 \affiliation{Instituto de F\'{i}sica, Universidade de S\~{a}o Paulo, Rua do Mat\~{a}o 1371, S\~{a}o Paulo 05508-090, Brazil}
 
\author{Yves Elskens}
 \affiliation{Aix-Marseille Universit\'{e}, CNRS, UMR 7345 PIIM, F-13397, Marseille cedex 13, France}

\email{yves.elskens@univ-amu.fr}
\email{ibere@if.usp.br} 
\email{matheus\_jean\_l@hotmail.com.br} 

\date{\today}

\begin{abstract}
A phenomenon of emergence of stability islands in phase-space is reported for two periodic potentials 
with tiling symmetries, one square and the other hexagonal, inspired by bidimensional Hamiltonian 
models of optical lattices.
The structures found, here termed as island myriads, resemble web-tori with notable fractality and 
arise at energy levels reaching that of unstable equilibria.
In general, the myriad is an arrangement of concentric island chains with properties relying on the 
translational and rotational symmetries of the potential functions. 
In the square system, orbits within the myriad come in isochronous pairs and can have different 
periodic closure, either returning to their initial position or jumping to identical sites in neighbor 
cells of the lattice, therefore impacting transport properties. 
As seen when compared to a more generic case, \textit{i.e.} the rectangular lattice, the breaking 
of square symmetry disrupts the myriad even for small deviations from its equilateral configuration. 
For the hexagonal case, the myriad was found but in attenuated form, mostly due to extra instabilities 
in the potential surface that prevent the stabilization of orbits forming the chains.
\end{abstract}


\maketitle

\begin{quotation}
A phenomenon of appearance of multi-oscillatory motion, here called island myriads, is reported 
for periodic potentials inspired by Hamiltonian models of two-dimensional optical lattices. 
The two types of potentials considered here have a periodic structure with square and hexagonal symmetries, 
allowing them to tile the plane completely. 
This periodic tiling aspect, and its connection to translational and rotational symmetries, is shown to be 
responsible for the existence of stable periodic orbits which will form the complex fractal-like structure that is 
the myriad. 
The effect of breaking these symmetries is also analyzed.
Besides, despite being related to the appearance of stable trajectories, the phenomenon takes place at 
energy values close to unstable equilibrium points of the potential surface.
\end{quotation}

\section{Introduction}\label{sec:introduction}

In a broad sense, among the various dynamical behaviors seen in Hamiltonian systems, the bifurcation 
of periodic orbits provides the main approach for describing changes in phase-space as one 
alters the control parameters of a model.
Particularly within area-preserving systems with more than one degree of freedom, bifurcations are 
extensively documented in the literature \cite{Contopoulos}. 
In this scenario, stable periodic orbits represent the elliptic centers of islands 
whereas when unstable they represent hyperbolic points, with their manifolds governing the dynamics 
within chaotic regions. 
As periodic solutions appear, disappear or change stability, they modify the kinetics of the system, 
thus being commonly referred to as the skeleton of phase-space dynamics.

These bifurcation processes are then applied in studies of the onset of chaos for resonant islands with 
commensurate frequency (in the context of KAM theorem \cite{Walker-Ford}); in the disruption of 
transport barriers, whether in twist or non-twist systems \cite{Castillo-Negrete}; to the existence 
of multiple oscillatory solutions (in the context of Birkhoff's theorem \cite{Reichl}); or diffusion 
\cite{Zaslavsky_2}, among others. 

A peculiar scenario is the one of web-tori, where a multitude of islands tiles a portion of 
phase-space, corresponding to a different bifurcation condition than that of KAM islands \cite{Zaslavsky}. 
In KAM theory, the persistence of invariant tori for a Hamiltonian $H_0(I)$ that is monotonic in the 
actions $I=(I_1, ..., I_n)$, is guaranteed for nonlinear perturbations with small enough amplitude. 
In the case of web-tori, the non-degeneracy condition 
is not satisfied, therewith allowing for the generation of multiple islands depending on the number of 
fixed points from the equations of motion and the frequency of the oscillatory perturbation.
Despite the good introduction to the topic given by Zaslavsky\cite{Zaslavsky_2} and 
Chernikov \cite{Chernikov}, not many works concern a detailed description of the properties of 
web-tori.
This configuration is mostly studied as a framework for weak chaos within stochastic webs, when 
separatrices are slightly perturbed forming thin chaotic channels with anomalous transport.

In this work, two Hamiltonian systems with periodic potentials, based on optical lattice models, are 
used to report a bifurcation phenomenon here named as island myriad. The myriad is a seemingly dense 
filling of a finite region of phase-space by a series of stability islands, highly 
resembling finite web-tori with notable fractality. It emerges from the bifurcation of orbits 
scattered when approaching a set of unstable equilibria with the same energy in potential surfaces with 
tiling symmetry. 
At the energy level of these equilibria, the deviated orbits form a fractal arrangement of isochronous 
island chains in phase-space, with each one being formed by rotated and translated 
symmetrical sets of orbits as allowed by the potential symmetries.
The two lattice models considered here have rectangular and hexagonal bidimensional tilings, and the myriad 
phenomenon is described in terms of its main periodic orbits and shown to rely on the symmetries of the 
potential functions.

Although the myriad was already reported in a previous work, the results presented here extend and 
are connected to the previous study done on the square lattice system \cite{Lazarotto}, to which we 
refer the reader for additional aspects of the dynamics regarding diffusion.

In what follows, section \ref{sec:model} starts with deducing the periodic potentials and further 
Hamiltonians for the selected systems, as inspired by a classical treatment of an optical lattice 
model.
The island myriad bifurcation is initially presented for the square lattice at section 
\ref{sec:results-square} and its properties listed for this case. 
Then, the effect of symmetry breaking is demonstrated by considering a non-equilateral 
tiling, \textit{i.e.} a generic rectangular lattice (sec. \ref{sec:results-rectangular}). 
At last, the myriad is further analyzed for the hexagonal system (sec. \ref{sec:results-hexagonal}). 
Appendices \ref{sec:append:single-coupling} to \ref{sec:append:triple-isochronicity} provide details 
for discussions made along the results sections.

\section{Lattice hamiltonian model}\label{sec:model}

Periodic potential models have been used as simple yet rich descriptions in different physical contexts, 
from cold-matter physics \cite{Bloch1}, charged particles in plasmas \cite{Kleva} or diffusion over 
crystal surfaces \cite{Sholl}.
The kind of periodic potential considered for this work is based on a classical Hamiltonian description 
of optical lattices. 

In these models, when a neutral atom interacts with an electric field $\vec{E}$ 
from a monochromatic 
wave, despite its neutrality, a dipole is induced along the field direction \cite{Grimm}.
The induced dipole thus oscillates with the wave while re-interacting with the field, hence submitting 
the particle to a potential function that, when averaged over time, results in a pondemorative 
potential 
\begin{equation}\label{eq:dipole-interaction}
\begin{split}
    V_\textrm{latt} &\approx - \frac{\rho(\omega)}{2} \lt| \vec{E}(\vec{r}) \rt|^2.
\end{split}
\end{equation}
Consequently, this conservative 
potential produces a force over the particle towards the wave antinode (node) if $\rho > 0$ ($\rho < 0$), 
thereby constraining the particle along the wave propagation axis.

From the single wave-particle interaction described, a generic lattice is then built by superposing 
the electric fields from multiple waves oriented throughout 3D space
\begin{equation}\label{eq:electric-field}
    \vec{E}_\textrm{latt}(\vec{r}) = \sum_{i=1}^N \hat{e}_n \; E_0^n \cos\lt(\vec{k}_n \cdot \vec{r} + \phi_n\rt) e^{-\mathrm{i} \omega_n t},
\end{equation}
with each one given by its polarization direction $\hat{e}_n$, amplitude $E_0^n$, wave vector $\vec{k}_n$, 
phase $\phi_n$ and angular frequency $\omega_n$. Note that $\hat{e}_n \cdot \vec{k}_n = 0 \;\forall n$.

For simplicity, all waves are assumed to have equal amplitude, resulting in the generic lattice potential
\begin{equation}\label{eq:latt-potential-generic}
\begin{split}
    V_\textrm{latt}\lt(\vec{r}\rt) = &\; U' \lt(\sum_{n=1}^N \cos^2 \lt(\vec{k}_n\cdot\vec{r}\rt) \rt. \\
                                            &\lt. + 2 \sum_{n=1}^N \sum_{m > n}^N \alpha_{nm} \cos \lt(\vec{k}_n\cdot\vec{r}\rt) 
                                                                                              \cos \lt(\vec{k}_m\cdot\vec{r}\rt)\rt)
\end{split}
\end{equation}
where $\alpha_{nm} = (\hat{e}_n \cdot \hat{e}_m) \cos(\phi_n - \phi_m)$ are the coupling parameters between waves $n$ and $m$ 
and $U' = -\frac{1}{2}\rho(\omega) E_0^2$.
In such a manner, the dynamics of particles is governed by $V_\textrm{latt}\lt(\vec{r}\rt)$ which generally is tridimensional.

In potential (\ref{eq:latt-potential-generic}), a variety of lattices can be built when combining different wave orientations and number. 
For 2D lattices, at least two co-planar, linearly independent wave vectors must be selected, producing a potential surface over the plane. 
To provide a $q$-fold symmetry to the lattice, one can set $q$ equally spaced wave vectors with the same norm
\begin{equation}\label{eq:wave-vectors}
    \vec{k}_n = k \cos\lt(n \frac{\pi}{q}\rt) \hat{x} + k \sin\lt(n \frac{\pi}{q}\rt) \hat{y},
\end{equation}
for $n = 0, ..., q-1$, generating polygonal lattice patterns.

Therefore, the Hamiltonian for a single particle in a bidimensional lattice is directly written as
\begin{equation}\label{eq:rectangular-hamiltonian}
\begin{split}
    H' = \frac{1}{2m} \lt(p_x^2 + p_y^2\rt) + V(x,y).
\end{split}
\end{equation}
for any lattice potential $V(x,y)$.

\subsection{Rectangular lattice}\label{sec:model:square-latt}

The main lattice type considered here will be a rectangular one, that is, the one formed by two 
perpendicular waves within the $x\!-\!y$ plane, where $\vec{k}_x = k_x\hat{x},\; \vec{k}_y = k_y\hat{y}$, 
yielding the periodic potential function 
\begin{equation}\label{eq:rectangular-potential}
\begin{split}
    V(x, y) = U' &\lt(\cos^2(k_x x) + \cos^2(k_y y) \;+\rt. \\
                 &\lt.\; 2 \alpha \cos(k_x x) \cos(k_y y)\rt),
\end{split}
\end{equation}
with 
\begin{equation}\label{eq:rectangular-parameters}
    U' = -\frac{1}{2}\rho(\omega)E_0^2 \quad\textrm{and}\quad \alpha = \lt(\hat{e}_x \cdot \hat{e}_y\rt) \cos(\phi_x - \phi_y).
\end{equation}

The resulting Hamiltonian (\ref{eq:rectangular-hamiltonian}) is re-scaled to $H = 2m H'$, 
so that the energy scale is $U=2m U'$. 
In the classical regime, the magnitude of $U$ has no relevance to the topology of solutions whatsoever, 
with only its sign being relevant. For this work, we consider the case of $U > 0$, in which the myriad 
could be identified. In case $U < 0$, the stability of equilibrium points is reversed and the dynamics 
is considerably different. 
Therefore, we set $U=20$ in agreement with Horsley \textit{et al.} \cite{Horsley}, 
although for simplicity it could be set to 1 without loss of generality. 

\begin{figure}[H]
    \centering
    \includegraphics[width=0.5\textwidth]{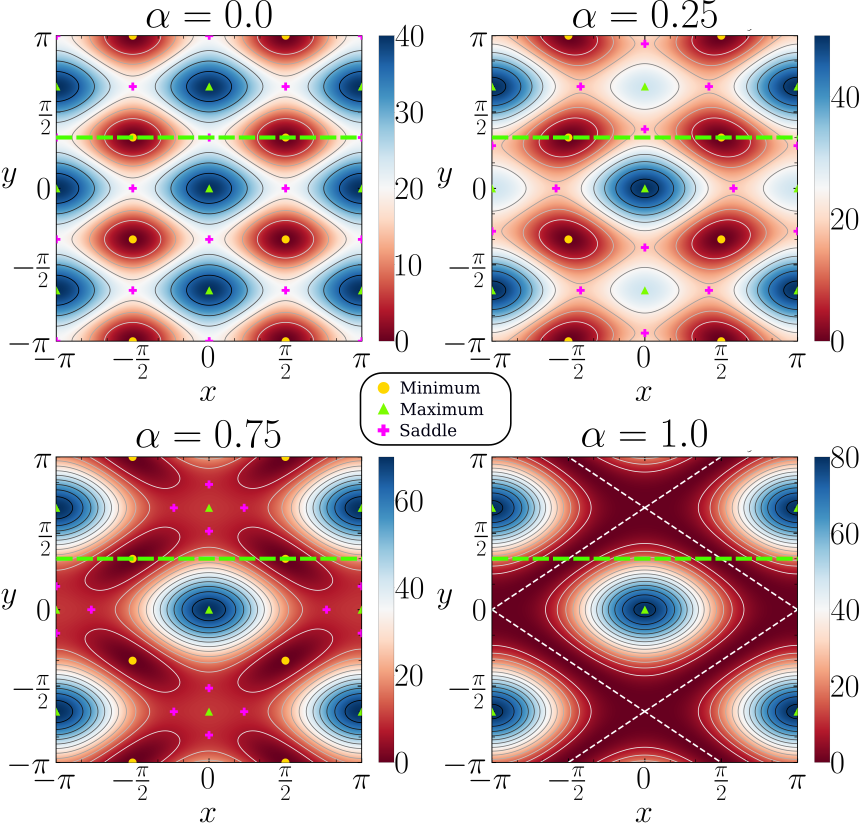}
    \caption{Color plot of the potential surface $V(x, y)$ (eq. \ref{eq:rectangular-potential}) 
             for $k_x=1, k_y=1.5$ and different values of the coupling $\alpha$. The Poincar\'{e} 
             section used for phase-space display is shown as the dashed green line at 
             $y = \frac{\pi}{2 k_y}$. At $\alpha = 1$, minima collapse into a `minima trench' given by 
             $\cos(k_y y) = -\cos(k_x x)$, shown as the white dashed line. 
             Color bars show the value of $V(x, y)$ in normalized units.}
    \label{fig:rectangular-lattice}
\end{figure}

The dynamics of a particle will then take place over the potential surface shown in figure 
\ref{fig:rectangular-lattice}, where it can be either trapped around minima regions, for 
energies below those of saddle points between potential wells, or otherwise wander to 
neighboring cells above this threshold. One may notice that a unit cell for the lattice 
can be defined as the box 
$(x,y) \in [-\frac{\pi}{k_x}, \frac{\pi}{k_x}] \times [-\frac{\pi}{k_y}, \frac{\pi}{k_y}]$, 
allowing for periodic boundary conditions when simulating trajectories 
(as used in this work when showing trajectories on the lattice unit cell). 
Notice that the near-symmetry case is when $k_x \approx k_y$, as the unit cell becomes square.

For increasing $\alpha$, the potential surface moves from the separable case ($\alpha = 0$) 
to fully superposed ($\alpha=1$) as the equilibrium points change energy and position (see fig. \ref{fig:rectangular-lattice} 
and table \ref{tab:rectangular-equilibrium-points}). While minima remain with zero energy and do not 
change position with varying $\alpha$, saddle points move towards local maxima, finally merging when $\alpha=1$, 
forming `trench lines' with degenerate minima along the lines $k_y y \pm k_x x \equiv \pi \mod(2 \pi)$. 
Simultaneously, local maxima diminish in energy thereby widening the pass between potential wells and 
facilitating the transport of particles. 

As seen in potential (\ref{eq:rectangular-potential}), the coupling parameter $\alpha$ acts as a 
perturbation to an integrable Hamiltonian of two uncoupled pendula-like potentials along $x$ and $y$ (with spatial 
period $\frac{2\pi}{k_i}$), coupling them for any $\alpha \neq 0$. Although $\alpha$ may vary in the interval $[-1, 1]$, 
one can limit oneself to solutions for $\alpha \in [0, 1]$ as the change $\alpha \to -\alpha$ is equivalent 
to a spatial translation by $\frac{\pi}{k_i}$ in one of the cartesian directions, thus not altering solutions properties.

For the purpose of this work, we start by considering the particular case of a square lattice, that is 
when $k_x=k_y$ (which is set as $k=1$ without loss of generality). In general, for any $k_x$, $k_y$, the rectangular 
lattice presents translational symmetry for displacements of $\frac{2 \pi}{k_i}$, for $i=x,y$, along each respective 
axis. However, the square lattice presents an extra rotation symmetry, by rotations of $\frac{\pi}{2}$. As will be 
discussed in section \ref{sec:results-square}, the presence of symmetry is necessary for the myriad existence. 
Considering this, we initially analyze the myriad for the square system and then set different $k_x$, $k_y$ such 
that symmetry is broken and its effects evaluated. Other alternatives for symmetry breaking 
are possible, such as the use of non-harmonic waves, as done by Porter \textit{et al.} \cite{Porter1}, 
although no particular analysis on the bifurcations of orbits is made in that work.

\begin{table}[H]
    \centering
    \begin{tabular}{l|c|c} \hline\hline 
    Equilibrium point          & $(x^*, y^*)$                                           & $V(x^*, y^*)$                    \\ \hline
\multirow{4}{*}{Minima}         & $\lt( \frac{\pi}{2 k_x},  \frac{\pi}{2 k_y}\rt)$       & \multirow{4}{*}{$0$}             \\
                                & $\lt(-\frac{\pi}{2 k_x}, -\frac{\pi}{2 k_y}\rt)$       &                                  \\
                                & $\lt( \frac{\pi}{2 k_x}, -\frac{\pi}{2 k_y}\rt)$       &                                  \\                                
                                & $\lt(-\frac{\pi}{2 k_x},  \frac{\pi}{2 k_y}\rt)$       &                                  \\ \hline
\multirow{2}{*}{Maxima (global)}& $(0, 0)$                                               & \multirow{2}{*}{$2U(1+\alpha)$}  \\ 
                                & $(\frac{\pi}{k_x}, \frac{\pi}{k_y})$                   &                                  \\ \hline 
\multirow{2}{*}{Maxima (local)} & $(\frac{\pi}{k_x}, 0)$                                 & \multirow{2}{*}{$2U(1-\alpha)$}  \\
                                & $(0, \frac{\pi}{k_y})$                                 &                                  \\ \hline
\multirow{4}{*}{Saddles}        & $(0, \pm\frac{1}{k_y}\cos^{-1}(-\alpha))$              & \multirow{4}{*}{$U(1-\alpha^2)$} \\ 
                                & $(\pm\frac{1}{k_x}\cos^{-1}(-\alpha), 0)$              &                                  \\ 
                                & $(\frac{\pi}{k_x}, \pm\frac{1}{k_y}\cos^{-1}(\alpha))$ &                                  \\
                                & $(\pm\frac{1}{k_x}\cos^{-1}(\alpha), \frac{\pi}{k_y})$ &                                  \\ \hline\hline
    \end{tabular}
    \caption{Equilibrium points position $(x^*, y^*)$ and energy $V(x^*, y^*)$ within 
             a unit cell of the rectangular lattice potential; positions are taken 
             modulo $2\pi/k_i$.}
    \label{tab:rectangular-equilibrium-points}
\end{table}

To display phase-space portraits, along all this work the Poincar\'{e} 
section with the oriented surface over two of the lattice minima
\begin{equation}\label{eq:pss}
    \Sigma = \lt\{\lt(x, y, p_x, p_y\rt) \in \mathbb{R}^4;\; y = \frac{\pi}{2 k_y};\; p_y > 0\rt\}
\end{equation}
will be used -- as highlighted in green in figure~\ref{fig:rectangular-lattice}.
Since Hamiltonian (\ref{eq:rectangular-hamiltonian}) is autonomous, energy ($E=H$) is an 
immediate constant of motion, constraining trajectories in a three-dimensional surface, which 
can thus be pictured in a 2D section. For the square lattice, the oriented surface $\Sigma$ is 
particularly convenient since its projection at the $(x,y)$ plane contains 
the minima at $x = \pm\frac{\pi}{2 k_x}$.
Indeed, bounded solutions around minima with $y < 0$ will occur, but nonetheless the $\frac{\pi}{2}$ 
rotation invariance implies that their symmetrical rotated counterpart solution will intersect 
$\Sigma$ at $y = \frac{\pi}{2}$.

\subsection{Hexagonal lattice}\label{sec:model:hexagonal-latt}

Analogous to rectangular lattices, hexagonal ones (or honeycomb lattices) are achieved with 3 co-planar 
wave vectors with same norm and equally spaced by $60^\textrm{o}$ from each other, in accordance to equation 
(\ref{eq:wave-vectors}). From equation (\ref{eq:latt-potential-generic}) the resulting potential is given by
\begin{widetext}
\begin{equation}\label{eq:honeycomb-potential}
\begin{split}
    V(x,y) = U' \Bigg(&\cos^2 (kx) + \cos^2\lt( \frac{k}{2} x + \frac{\sqrt{3} k}{2} y\rt) + \cos^2\lt(-\frac{k}{2} x + \frac{\sqrt{3} k}{2} y\rt) 
                       + 2 \alpha_{12} \cos\lt(k x\rt) \cos\lt(\frac{k}{2} x + \frac{\sqrt{3} k}{2} y\rt) \\
                      &+ 2 \alpha_{13} \cos\lt(k x\rt) \cos\lt(-\frac{k}{2} x + \frac{\sqrt{3} k}{2} y\rt)
                       + 2 \alpha_{23} \cos\lt(\frac{k}{2} x + \frac{\sqrt{3} k}{2} y\rt) \cos\lt(-\frac{k}{2} x + \frac{\sqrt{3} k}{2} y\rt) \Bigg).
\end{split}
\end{equation}
\end{widetext}

\begin{figure}[H]
  \centering
  \includegraphics[width=0.5\textwidth]{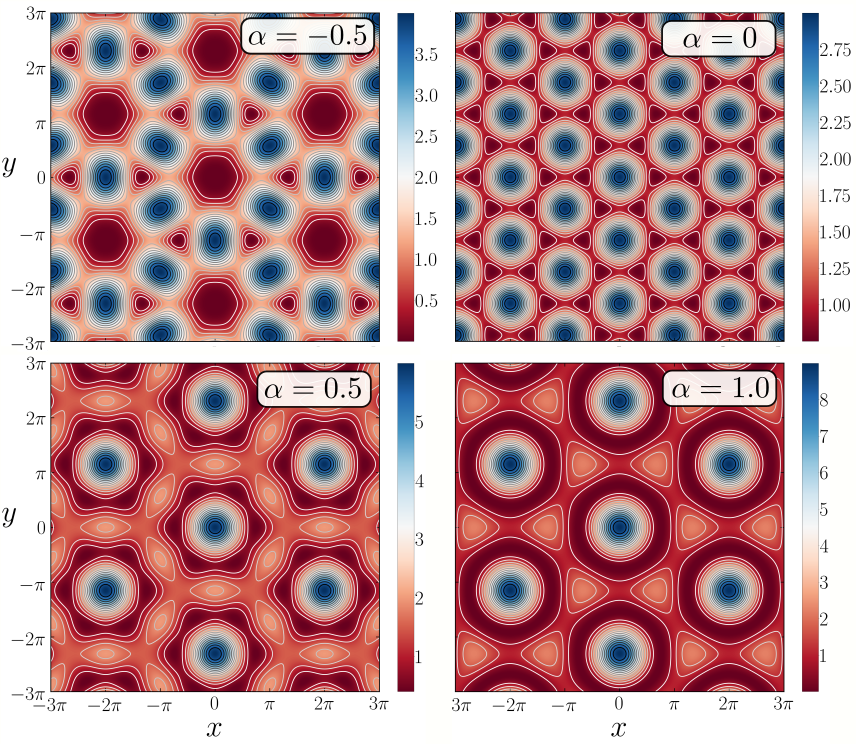}
  \caption{Color plot of the hexagonal lattice potential (\ref{eq:honeycomb-potential}) for 
           different $\alpha$ values, within the single parameter condition ($\alpha_{nm} = \alpha$). 
           For these figures $U=1$. Color bars show the value of $V(x, y)$ in normalized units.}
  \label{fig:honeycomb-potential-plot}
\end{figure}

\begin{figure}[h]
    \includegraphics[width=0.5\textwidth]{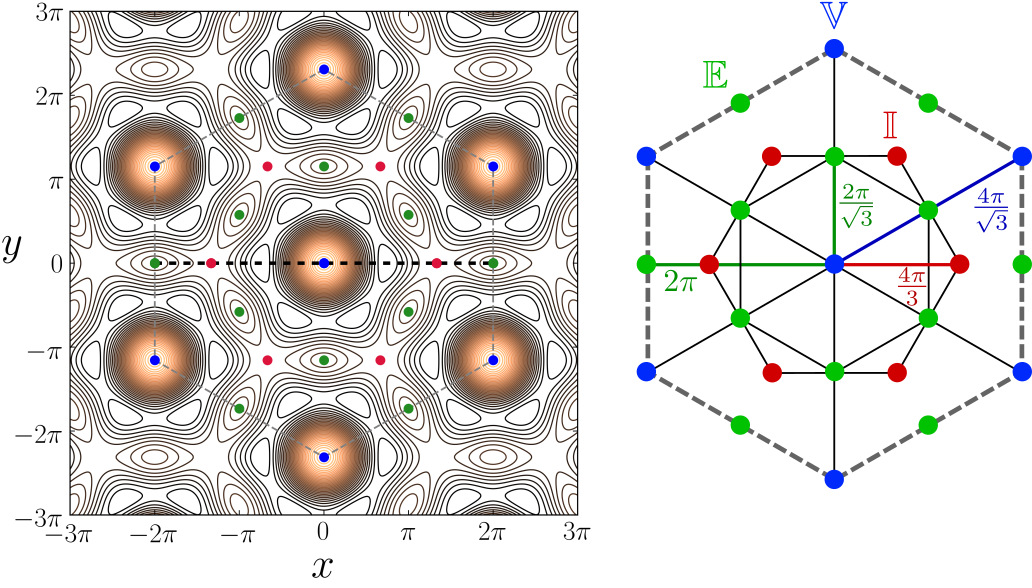}
    \caption{Geometric schematic used for equilibrium points calculation in the hexagonal lattice 
             potential, assuming the single coupling condition. (Left) Equipotential lines (for 
             $\alpha = 0.5$) and the selected equilibria. The black dashed line marks the Poincar\'{e} section 
             position at $y = 0$. (Right) Schematization of the unit cell with the main hexagons 
             and distances used for calculations. Equal colors indicate equal energy levels.}
    \label{fig:honeycomb-equilibrium-points}
\end{figure}

The potential form (\ref{eq:honeycomb-potential}) has three coupling parameters $\alpha_{nm}$, 
implying a 4-dimensional parameter space: ($E, \alpha_{12}, \alpha_{13}, \alpha_{23}$); 
it is then convenient to reduce it. 
For this purpose, figure \ref{fig:honeycomb-potential-plot} shows potential surfaces for different 
values of $\alpha$, notably for cases where all coefficients are equal ($\alpha_{12} = 
\alpha_{13} = \alpha_{23} = \alpha$).
With this condition, some equilibria alter their energy value and stability while remaining 
in a regular hexagonal structure (fig. \ref{fig:honeycomb-equilibrium-points}). 
Hence, as $\alpha$ varies, the potential surface changes in a similar fashion to that seen 
for the square lattice, where points may change their energy or stability, while keeping their symmetrical 
positions with fixed distances (fig. \ref{fig:honeycomb-equilibrium-points}, right frame). 
Although restrictive, this simplification ensures the parameter space reduction 
and the preservation of symmetry required for the purposes of this study, as will be made clear when 
discussing the dependence of the myriad phenomenon with the latter aspect.

Nonetheless, since the orientation of one wave-vector alters its coupling with all the 
other waves, it can be shown that when all $\alpha_{nm}$ are equal, they must lie in the range 
$\alpha \in [-\hlf, 1]$ (for details, see appendix \ref{sec:append:single-coupling}).

In this single coupling scenario, the potential can be re-written as 
\begin{eqnarray}\label{eq:honeycomb-potential-single-coupling}
  V(x,y) &=& U \Biggl(1 + \lt(\alpha + \cos(x)\rt) \cos\lt(\sqrt{3} y\rt) \Biggr. \nonumber \\
         && \quad + \cos(x) \Bigl( 4 \alpha \cos\lt(\frac{x}{2}\rt) \cos\Bigl(\frac{\sqrt{3} y}{2} \Bigr) \Bigr. \nonumber \\
         && \qquad \qquad \qquad \Biggl. \Bigl. + \alpha + \cos(x)\Bigr)\Biggr),
\end{eqnarray}
and consequently the Hamiltonian is as in equation (\ref{eq:rectangular-hamiltonian}) with 
$V(x,y)$ either as in equation (\ref{eq:honeycomb-potential}) or (\ref{eq:honeycomb-potential-single-coupling})
and $H$ normalized as in the rectangular lattice.
Similar to the rectangular system, for the hexagonal lattice, phase-space displays will be 
made over the section placed at $y=0$ and oriented as $p_y > 0$ (dashed black line in the left 
frame of fig. \ref{fig:honeycomb-equilibrium-points}).

\begin{table}[H]
  \begin{tabular}{c|c|c} \hline\hline
\rule{0pt}{3ex}    Eq. point                    & $(x^*, y^*)$                                & $V(x^*, y^*)$            \\ \hline
\multirow{2}{*}{\tikzcircle[blue, fill=blue]{2.0pt} $\mathbb{V}$} & $(0,\;0)$                          & \multirow{2}{*}{$3 \lt(1 + 2 \alpha\rt)$} \\ 
                                                       & $\frac{4\pi}{\sqrt{3}}
                                                          \lt(\cos\lt(\frac{(2n+1)\pi}{6}\rt),\; 
                                                              \sin\lt(\frac{(2n+1)\pi}{6}\rt)\rt)$   &                                  \\ \hline
\multirow{2}{*}{\tikzcircle[ForestGreen, fill=ForestGreen]{2.0pt} $\mathbb{E}$} & $\frac{2\pi}{\sqrt{3}}
                                                                        \lt(\cos\lt(\frac{(2n+1)\pi}{6}\rt),\;  
                                                                        \sin\lt(\frac{(2n+1)\pi}{6}\rt)\rt)$ & \multirow{2}{*}{$3 - 2 \alpha$} \\
                                                                     & $2 \pi \lt(\cos\lt(\frac{n \pi}{3}\rt),\; 
                                                                                  \sin\lt(\frac{n \pi}{3}\rt)\rt)$ &           \\ \hline
\tikzcircle[BrickRed, fill=BrickRed]{2.0pt} $\mathbb{I}$       & $\frac{4\pi}{3}
                                              \lt(\cos\lt(\frac{n \pi}{3}\rt),\;  
                                                  \sin\lt(\frac{n \pi}{3}\rt)\rt)$      & $\frac{3}{4} + \frac{3\alpha}{2}$\vspace*{0.1cm} \\ \hline\hline
  \end{tabular}
  \caption{Equilibrium points position $(x^*, y^*)$ and energy $V(x^*, y^*)$ for some reference 
           points in the hexagonal unit cell (fig. \ref{fig:honeycomb-equilibrium-points}).
           In the indices above, $n=0,1,2,3,4,5$, modulo $6$.
           Information on the labels is given in the text.}
  \label{tab:honeycomb-equilibrium-points}
\end{table}

In the right frame of figure \ref{fig:honeycomb-equilibrium-points}, selected equilibria
are highlighted over the hexagonal lattice unit cell. They were selected both as geometrical 
and energetical references related to the expectation to find the island myriad phenomenon.
The unit cell vertices and center point are labeled as $\mathbb{V}$; the cell outermost edges 
and the inner hexagon edges are labeled as $\mathbb{E}$ and the innermost hexagon vertices as $\mathbb{I}$ 
(see table \ref{tab:honeycomb-equilibrium-points}).

\section{The island myriad -- square lattice}\label{sec:results-square}

We start by presenting the island myriad phenomenon for the square lattice in the context of  
emergence of stability structures in phase-space. 
When measuring the area (or volume) of phase-space occupied by islands or chaotic regions 
as a function of the control parameters $(E, \alpha)$, a series of fluctuations are expected given the 
mixed nature of nonlinear dynamics.
For this purpose, the chaotic/regular areas were measured over the section $\Sigma$ (eq. \ref{eq:pss}) via 
a smaller alignment index (SALI) method, as developed by Skokos\cite{Skokos_paper,Skokos_book}.

Briefly, the algorithm integrates a single orbit along with two deviation vectors ($\vec{\omega}_1(t), \vec{\omega}_2(t)$).
These vectors are evolved in time by the linearized equations of motion (and rescaled when necessary) and 
present different behavior depending on the nature of the orbit.
In case it is chaotic, the deviation vectors align or anti-align to each other due to the exponential 
stretching of phase-space along the unstable manifold direction. 
On the other hand, if the orbit is regular, ($\vec{\omega}_1, \vec{\omega}_2$) are kept at finite 
angle (up to secular drift) while only orienting themselves towards the tangent plane of the stable 
torus in which the orbit is contained.
Therefore, the evaluation of this alignment, achieved by the index function
\begin{equation}\label{eq:sali}
    \textrm{SALI}(t) \coloneqq \textrm{min}\lt(\lVert\hat{\omega}_1(t) + \hat{\omega}_2(t)\rVert, 
                                               \lVert\hat{\omega}_1(t) - \hat{\omega}_2(t)\rVert\rt),
\end{equation}
can numerically discriminate the orbit's stability, such that SALI($t$) $\to0$ exponentially as $t\to\infty$ for 
chaotic orbits, while it keeps an essentially constant non-zero value for regular ones 
(SALI$(t)\in(0,\sqrt{2}]$ -- for normalized $\hat{\omega}_i = \vec{\omega}_i / |\vec{\omega}_i|$)).
It is possible that for regular orbits the tangent vectors still align/anti-align due to shear between
close torus layers; however, this was seen to occur over times much longer than the one for alignment in 
chaotic orbits.

\begin{figure}[H]
    \centering
    \includegraphics[width=0.5\textwidth]{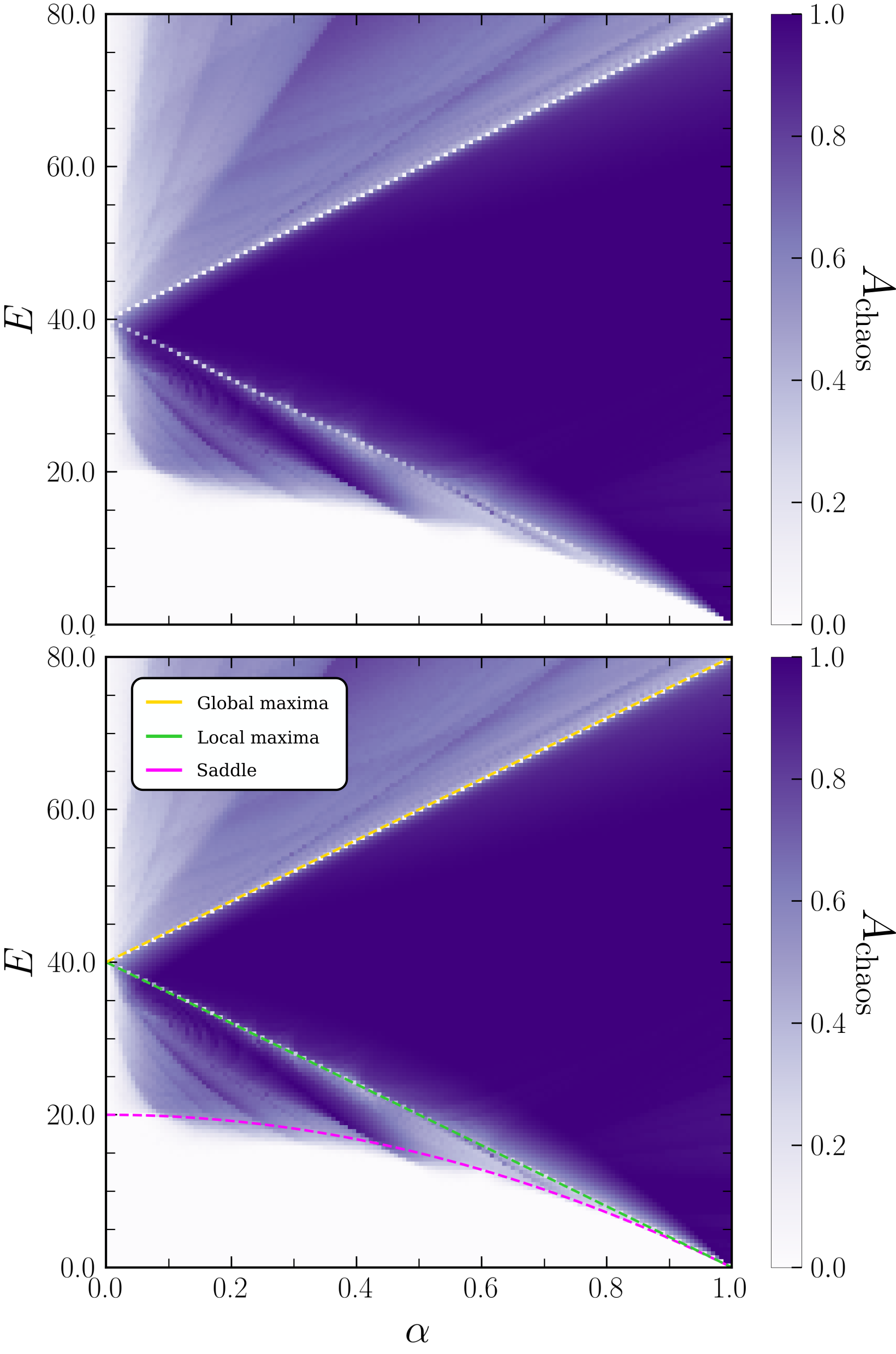}
    \caption{Color map of the chaotic area portion in parameter space for the square lattice. Total 
             chaos is indicated by $A=1$ and total regularity by $A=0$. In the top frame, 
             the seemingly dotted white lines are an artifact of grid pixel precision. 
             In the bottom frame, energy 
             lines for equilibrium points are displayed as: $V_\textrm{saddle} = U (1-\alpha^2)$ in 
             magenta; $V_\textrm{l-max} = 2U(1-\alpha)$ in green and $V_\textrm{g-max} = 2U(1 + \alpha)$ 
             in yellow. Grid size is $250 \times 250$.}
    \label{fig:chaotic-area-param-square-latt}
\end{figure}

Using such a discrimination index, the chaotic/regular areas are then identified over a 2D fine mesh of the surface 
$\Sigma$, where each area tile is attributed to an initial condition and all tiles are summed at the end. 
Figure \ref{fig:chaotic-area-param-square-latt} shows a color map of the chaotic area percentage ($A$) of phase-space 
along all parameter space ($E,\alpha$), revealing a series of patterns of emergence and disappearance of stability structures 
(white regions -- $A \approx 0.0$).
As seen in the bottom frame, two particular lines stand out with the dominance of stability structures as well as borders 
to the global chaos limit of the system ($A \approx 1.0$). These straight pixelated lines with $A \approx 0$ 
are seen to coincide with the energy levels $E(\alpha)$ of maxima of the square lattice potential, being 
negatively (positively) inclined for local (global) maxima (eq. \ref{eq:rectangular-potential} and 
table \ref{tab:rectangular-equilibrium-points}), as highlighted in the lower frame.

\begin{figure*}
    \centering
    \includegraphics[width=1.0\textwidth]{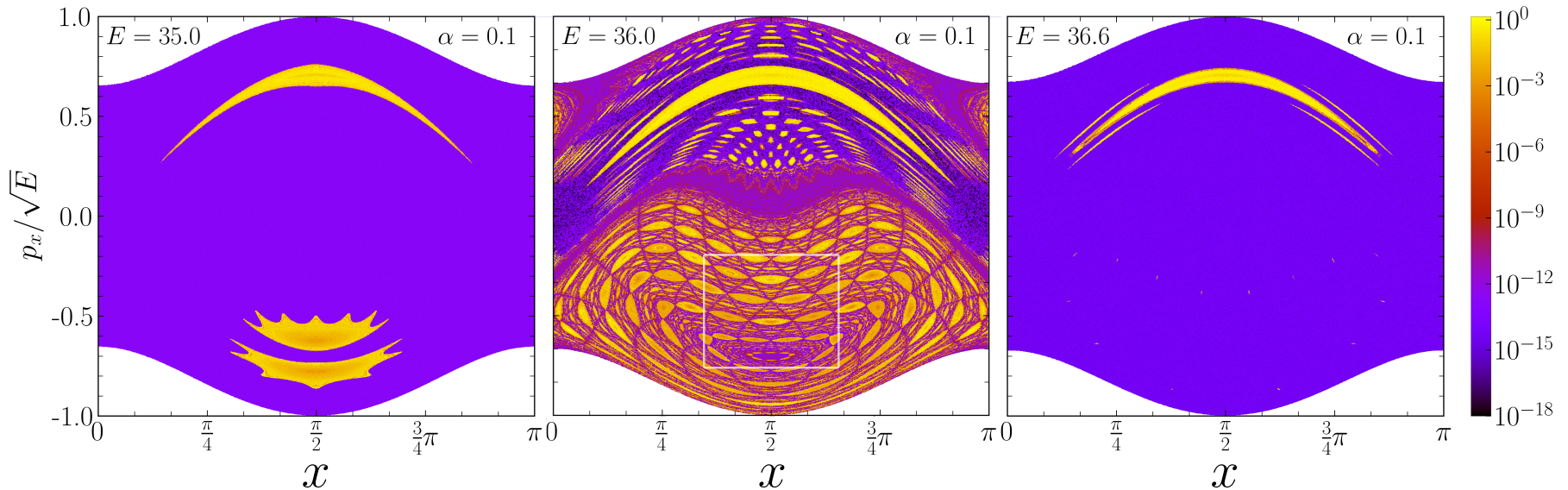}
    \caption{Colorized section $\Sigma$ for varying energy $E$. 
             (Left) $E=35$, below local maxima energy.
             (Center) Island myriad at $E = 36$, exactly at local maxima level: $V_\textrm{local}(\alpha=0.1)$ 
             (table \ref{tab:rectangular-equilibrium-points}) -- the white square indicates 
             a zoom in shown in figure \ref{fig:myriad-square-lattice}. 
             (Right) $E=36.6$, after the myriad disappearance. 
             Colors were set using the SALI$(t)$ index value to emphasize islands (in yellow) 
             from chaos (in purple).}
    \label{fig:myriad-transition}
\end{figure*}

In the light of this, it is seen that when the system energy reaches that of unstable equilibria, either local or global, 
stability structures emerge in phase-space. These structures are a myriad of island chains, as illustrated in figure 
\ref{fig:myriad-transition} for the case of $\alpha=0.1$ and $E=36$ (over the local maxima energy line -- in green in figure 
\ref{fig:chaotic-area-param-square-latt}).
Below local maxima energy, phase-space is dominated by a chaotic sea with three main stability islands. As energy 
approaches the local maximum level, the two bottommost islands vanish and the chaotic sea is filled with a multitude of island 
chains. 
Right above the local-maxima level ($E \approx 36.6$), the myriad completely vanishes and phase-space is again dominated by 
a uniform chaotic sea.

\begin{figure*}
    \centering
    \includegraphics[width=1.0\textwidth]{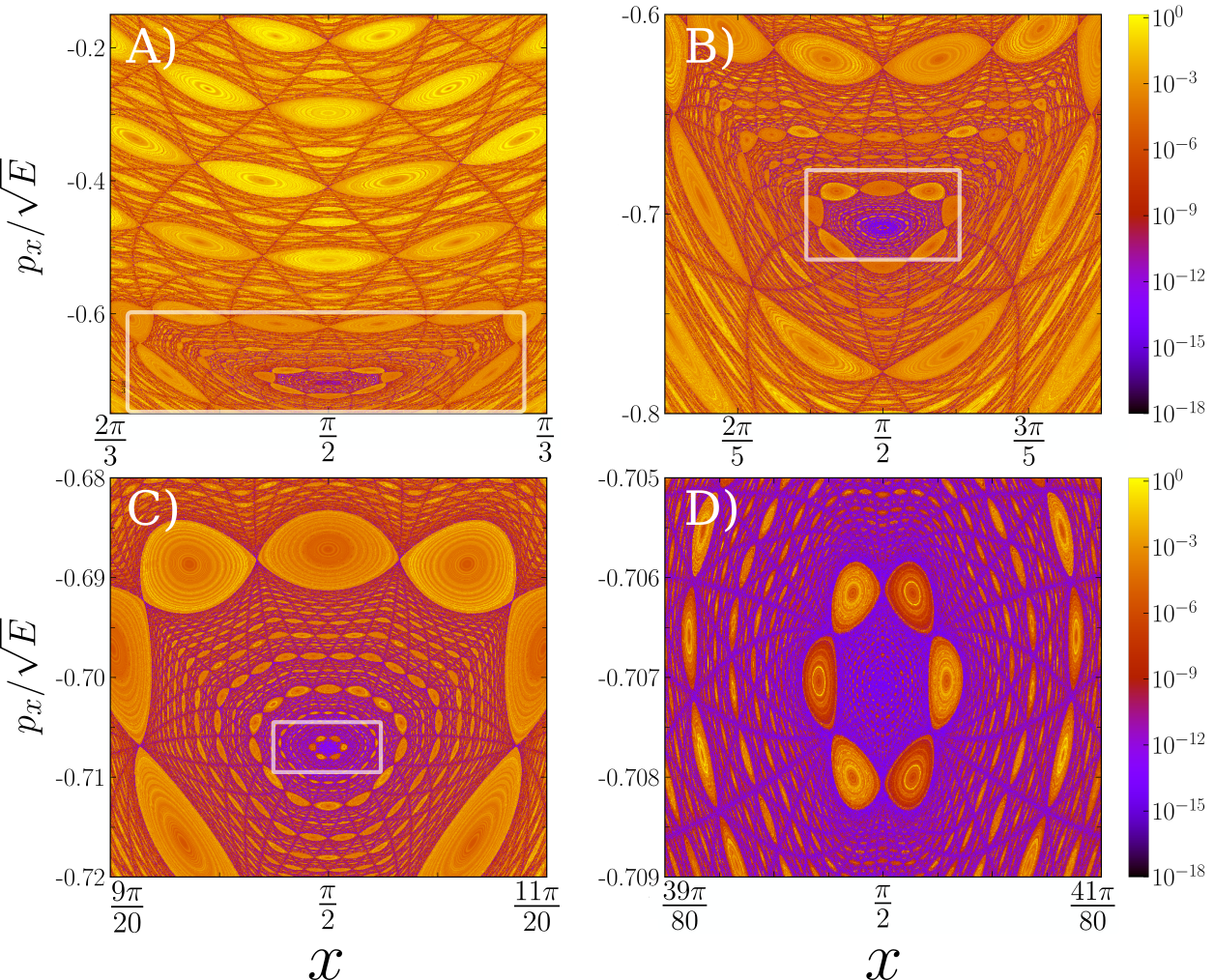}
    \caption{Colorized phase-space of the island myriad for $\alpha = 0.1$ and $E=36$ showing 
             successive zooms into the myriad core. 
             A) zoom in from the central frame of figure \ref{fig:myriad-transition}. 
             B) zoom in from frame A. 
             C) zoom in from frame B. 
             D) zoom in from frame C.}
    \label{fig:myriad-square-lattice}
\end{figure*}

The observed chains have always even period and are all concentric around the hyperbolic fixed point located at 
$(x, \frac{p_x}{\sqrt{E}}) \approx (\frac{\pi}{2}, -0.71$), forming an onion-like structure with apparent fractality, 
as higher period chains appear in between smaller period ones (fig. \ref{fig:myriad-square-lattice}). 
The mentioned hyperbolic point corresponds to the unstable periodic orbit located along the local maxima at 
$(x_\textrm{loc}, y_\textrm{loc}) = (0,\pi)$ and 
$(x_\textrm{loc}, y_\textrm{loc}) = (\pi,0)$. 
The myriad is more clearly visible in parameter space for $\alpha \lesssim 0.6$ and inside a short energy window of 
$\Delta E \approx 0.5$ above maxima energy values. 

As mentioned earlier, a similar stability emergence is seen around global maxima energy lines (yellow 
line in figure \ref{fig:chaotic-area-param-square-latt}). The myriad structure for this scenario is 
qualitatively similar to the one seen over local maxima lines, as shown in appendix 
\ref{sec:append:myriad-square-global-max}, so it will not be detailed in this work, which mainly 
focuses on case of local maxima energy levels. 

\subsection{Isochronicity}

\begin{figure*}
    \centering
    \includegraphics[width=1.0\textwidth]{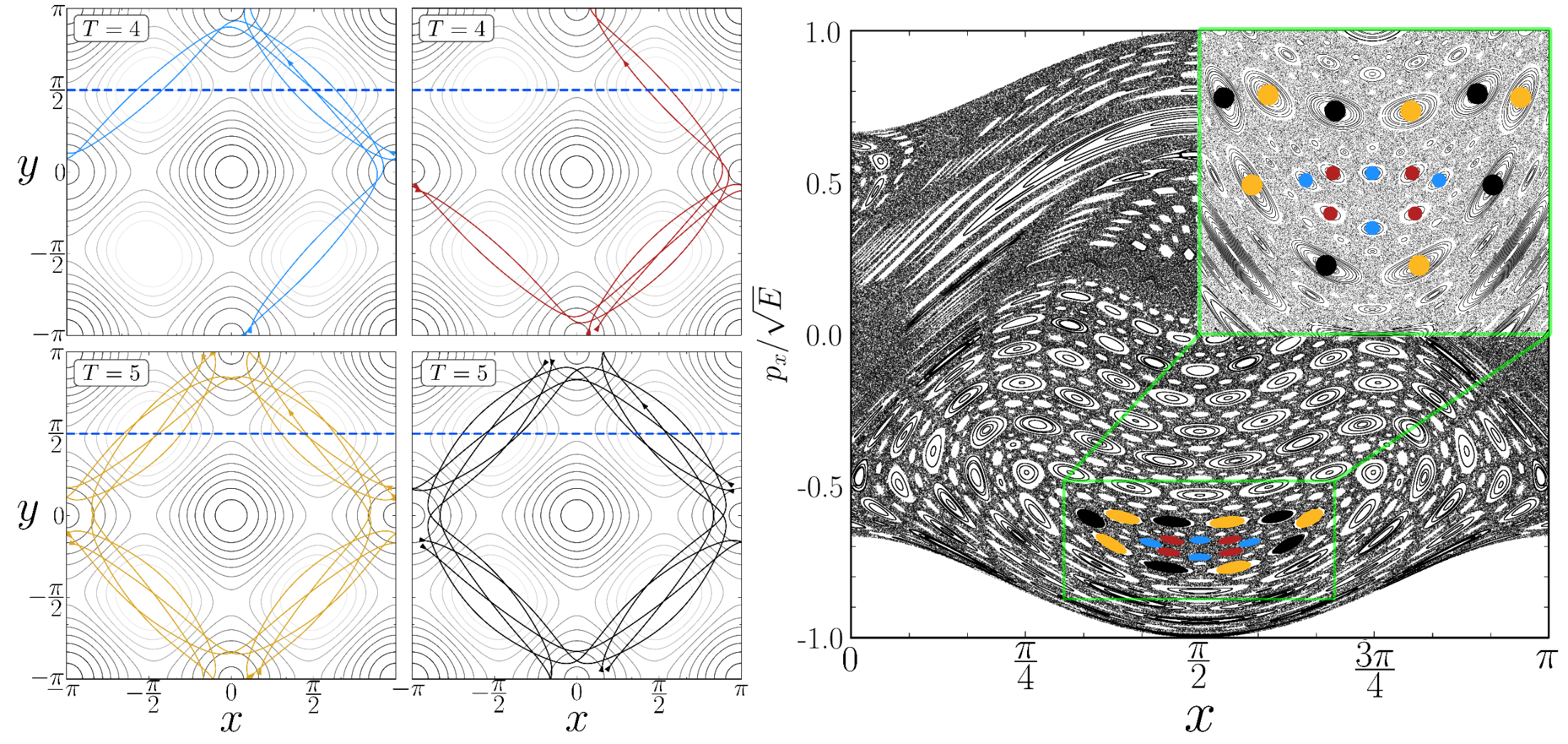}
    \caption{(Left) Isochronous periodic orbits from two island myriad chains, shown in the square lattice unit cell. 
             (Right) Poincar\'{e} section with the myriad structure. 
             Colored dots indicate the elliptic fixed points of the orbits relative to the section $\Sigma$ (blue dotted line); 
             $T$ indicates the fixed point period. The innermost period 8 chain is formed by the blue and red sets, while 
             the immediate next one, with period 10, by the yellow and black ones.}
    \label{fig:isochronous-chains}
\end{figure*}

An individual island chain is not related to a single stable periodic orbit and its set of 
elliptic fixed points, as one usually expects. Instead, all chains in the myriad are isochronous, 
in the sense that they are formed by two (or more) independent sets of interleaved island links, where 
orbits contained in one set do not overlap with the other, as illustrated in figure \ref{fig:isochronous-chains}.

The isochronous condition can be found in many dynamical systems \cite{Sousa, Bruno} but here its origin is clearly 
seen as a consequence of the system symmetries. As an example, figure \ref{fig:isochronous-chains} shows that the orbits forming each chain 
set are symmetric pairs (blue and red orbits for the period 8 chain, and yellow and black orbits for the period 10), 
\textit{i.e.} they are rotated by $(l+1) \frac{\pi}{2}$ or translated by ($n\pi, m\pi)$ in space, for $l, m, n\in\mathbb{Z}$, 
relative to each other. 
Since they are the same geometrical curve, they present the same period and rotation number, therefore 
emerging in the same torus layer. Indeed, this is an immediate consequence of the square lattice 
potential translational and rotational symmetries, as well as its `tiling' closure property, the same 
allowing for the use of periodic boundary conditions. In general, chains in the myriad present double 
isochronicity, being split into two orbit sets; however, triple isochronicity can also be found, as 
better detailed in appendix \ref{sec:append:triple-isochronicity}.

\subsection{Escape time (periodic spatial closure)}

\begin{figure}
    \centering
    \includegraphics[width=0.5\textwidth]{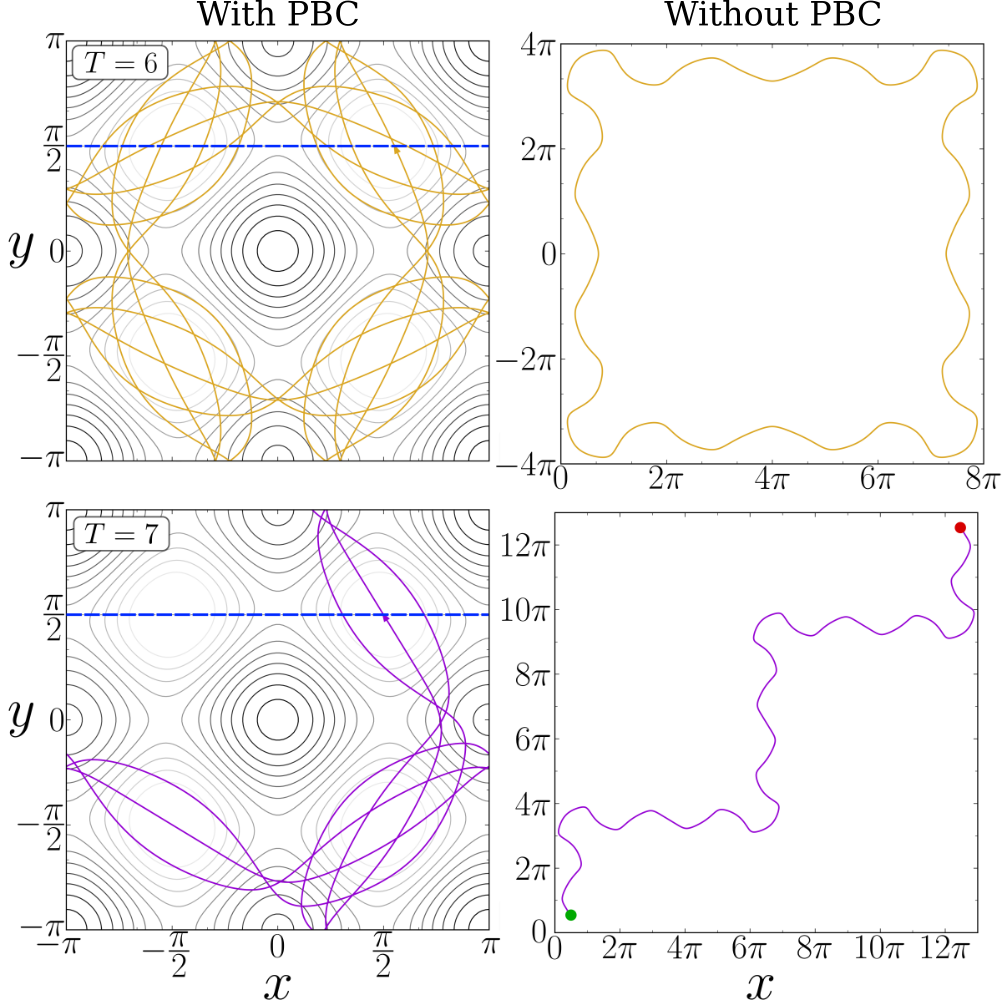}
    \caption{Examples of periodic orbits with different spatial closures. On the left column, trajectories within the 
             unit cell and boundary conditions applied. On the right column, the same trajectories without boundary conditions, 
             ranging through all space.}
    \label{fig:orbits-closure}
\end{figure}

When inspecting the stable periodic orbits associated with different chain layers in the myriad, 
distinct periodic behaviors are seen. 
In one case, orbits return to their exact initial position even when disregarding periodic boundary conditions (as in a libration), 
whereas in the other they only do so with them (as in a rotation), as exemplified in figure \ref{fig:orbits-closure}. 
This difference in spatial closure therefore directly impacts the transport properties between different myriad layers. 
In this context, escape time basins are simply defined as a color map of the time required for initial conditions on the section 
$\Sigma$ (within a central unit cell) to reach outside the square box with $n$ unit cells of size, 
\textit{i.e.}, $x,y \in \lt[-n\pi, n\pi\rt]$ (here $n = 20$).

\begin{figure}
    \centering
    \includegraphics[width=0.5\textwidth]{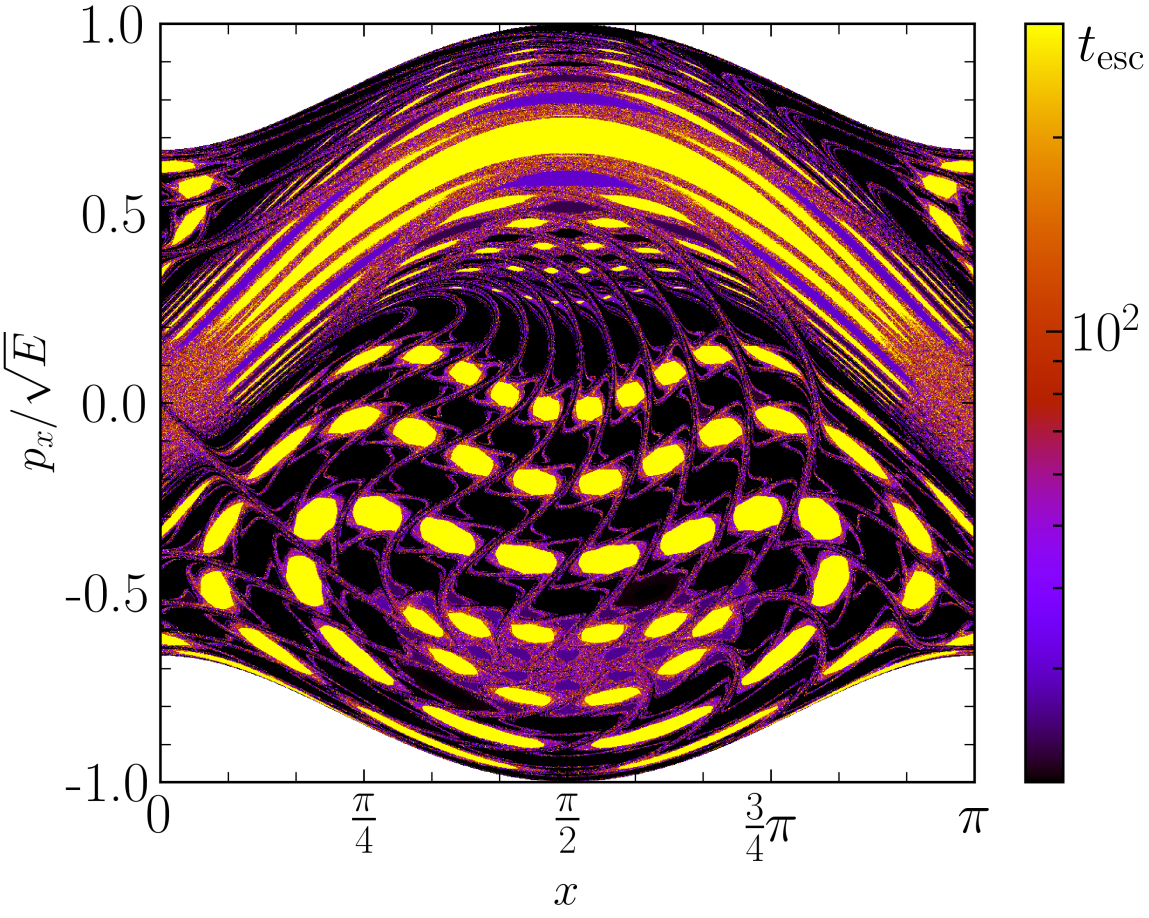}
    \caption{Color map of escape time ($t_\textrm{esc}$) basins over the island myriad. System parameters are $\alpha=0.1$, $E=36.05$.}
    \label{fig:square-lattice-escape-time}
\end{figure}

As seen among the interleaved chains, in yellow escape time basins in figure \ref{fig:square-lattice-escape-time}, 
orbits from islands with librational movement (top frames in fig. \ref{fig:orbits-closure}) 
remain trapped while the ones from purple basins with rotational movement (bottom frames in fig. \ref{fig:orbits-closure}) 
quickly escape in direct flights through the lattice. 
This is only possible due to the periodic `tiling' property of the potential function,  
as translated positions $(x, y) \to (x \pm 2n\pi, y \pm 2m\pi)$, for $n, m \in \mathbb{Z}$, will correspond 
to an identical site in a neighbor unit cell, thus allowing for periodic behavior without return to 
the exact initial position.

\subsection{Separatrix reconnection}

As asserted initially, the island myriad is expected to emerge when orbits reach the energy level 
of unstable equilibria of the lattice.
However, the energy of these points themselves changes with the coupling $\alpha$, thereby raising the 
possibility of analysing the myriad evolution as the unstable points change. 

Qualitatively, it was found that when varying the energy over the local maxima line for increasing coupling, 
\textit{i.e.} $E = V_\textrm{l-max} = 2 U (1 - \alpha)$ (in green in fig. \ref{fig:chaotic-area-param-square-latt}), 
all island chains move outwards from the myriad center; simultaneously, islands external to the myriad core, from 
an external layer surrounding the myriad, move inwards, eventually `colliding' with the outgoing inner island chains.

\begin{figure}
    \centering
    \includegraphics[width=0.5\textwidth]{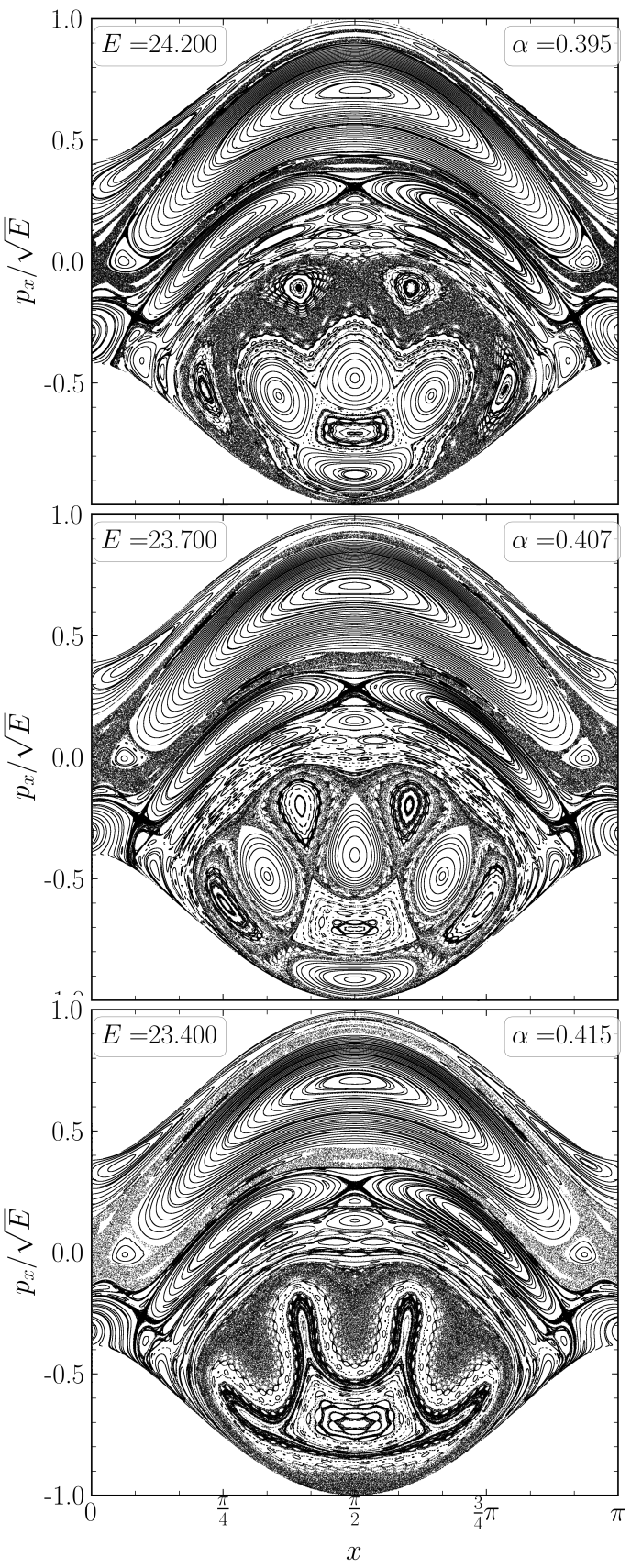}
    \caption{Separatrix reconnection of islands of period 4 within the myriad as ($\alpha$, $E$) 
             vary on the local maxima line in the square lattice. The coupling increases from top 
             to bottom, $\alpha = 0.295 \to 0.305$ and $E = V_\textrm{l-max}(\alpha)$.}
    \label{fig:reconnection_1}
\end{figure}

\begin{figure}
    \centering
    \includegraphics[width=0.5\textwidth]{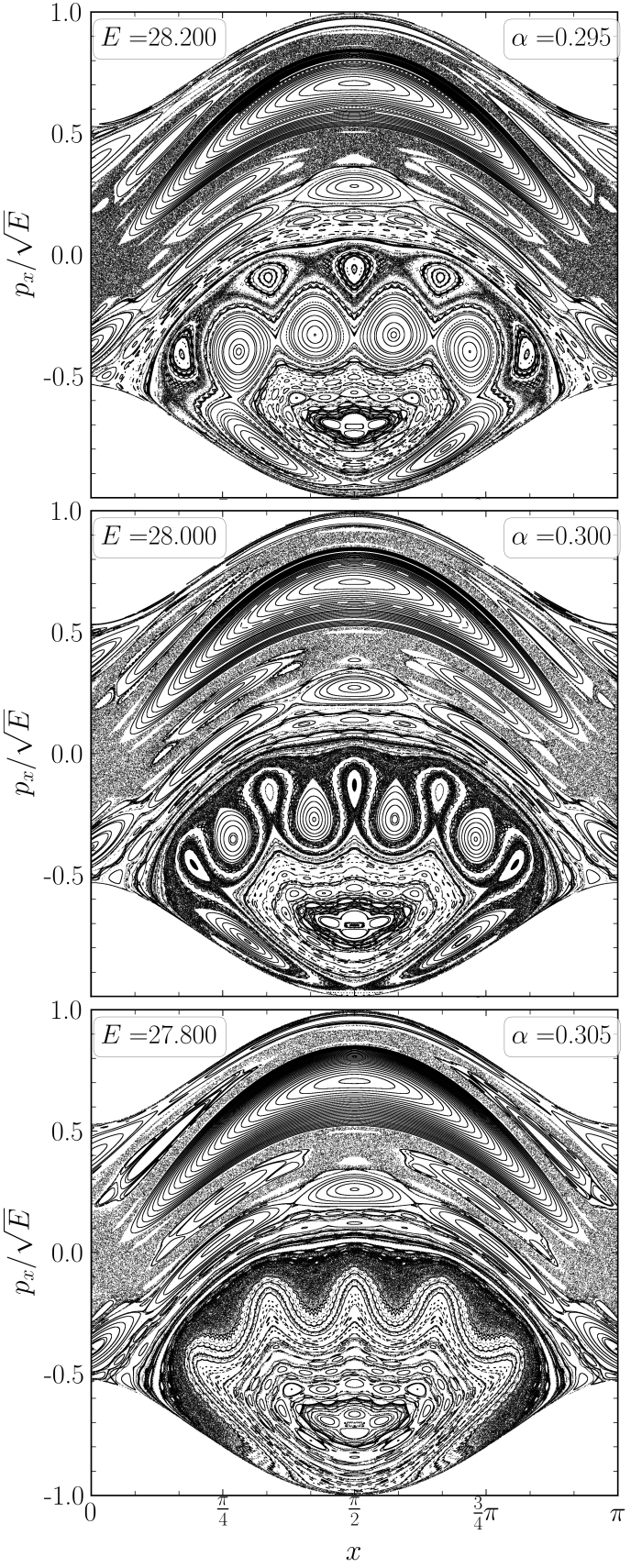}
    \caption{Separatrix reconnection of islands of period 6 within the myriad as ($\alpha$, $E$) vary on the 
             local maxima line in the square lattice. The coupling increases from top to bottom, 
             $\alpha = 0.395 \to 0.415$ and $E = V_\textrm{l-max}(\alpha)$.}
    \label{fig:reconnection_2}
\end{figure}

The `collision' (or superposition) of these island chains gives rise to a bifurcation process eventually leading 
to the disappearance of both chains. Particularly, this bifurcation process occurs via a separatrix reconnection, 
as illustrated for a pair of chains of period 4, in figure \ref{fig:reconnection_1}, and a pair of period 6, in figure 
\ref{fig:reconnection_2}.
In this process, the outer and inner chains are interdigitated relative to each other, in the sense that 
the stable centers of islands from one chain align with the saddles (unstable points) of the other. 
When colliding, the separatrix is divided while changing its configuration, with the previous 
outermost chain now inside the center myriad structure and the former inner chain immersed in the 
chaotic area. 
This process keeps on going continuously and sequentially as the inner chains move outwards, always 
in an interdigitated configuration relative to the outer ones, then reconnecting and further disappearing, 
eroding the myriad with chaos until it vanishes for $\alpha = 1$ and $E = 0$.

Commonly, the scenario of separatrix reconnection is seen in non-twist systems, widely studied in 
their standard form \cite{Castillo-Negrete, Morrison}. In such systems, the twist property, \textit{i.e.} 
the monotonic increase of the winding number with the action variable, is violated, presenting points 
of maximum or minimum. 
In case resonances appear around these extreme points, they form an interdigitated 
island chain pair, similar to that seen in figures \ref{fig:reconnection_1} and \ref{fig:reconnection_2}. At the same time, 
the curve between them, exactly at the extreme point, is a shearless curve which acts as a transport barrier 
between chaotic regions in phase-space.
Here, a similar arrangement is seen when considering the local winding number relative to the island 
myriad center. The supposed shearless curve would thus be expected to occur between the interdigitated 
islands that reconnect; however, higher order bifurcations and the constant presence of a chaotic layer 
between them prevents a direct verification via winding number profile 
and can indicate that the curve is destroyed.

\section{Island myriad -- rectangular lattice}\label{sec:results-rectangular}

The results obtained for the square lattice highlight the dependence of the myriad phenomenon 
on the tiling symmetry of the potential function, which stands from the assumption $k_x=k_y$. 
For this purpose, this section verifies to what extent the breaking of this symmetry affects the 
phenomenon.

When setting $k_x \neq k_y$, the rotation symmetry is lost, although translation symmetry is still 
preserved for translations by $\frac{2\pi}{k_i}$, for $i=x,y$, along the axis. Nevertheless, the stable periodic orbits 
that form the myriad observed in the square system may be deformed or change their stability as symmetry 
is broken, preventing its emergence.

\begin{figure}
    \centering
    \includegraphics[width=0.5\textwidth]{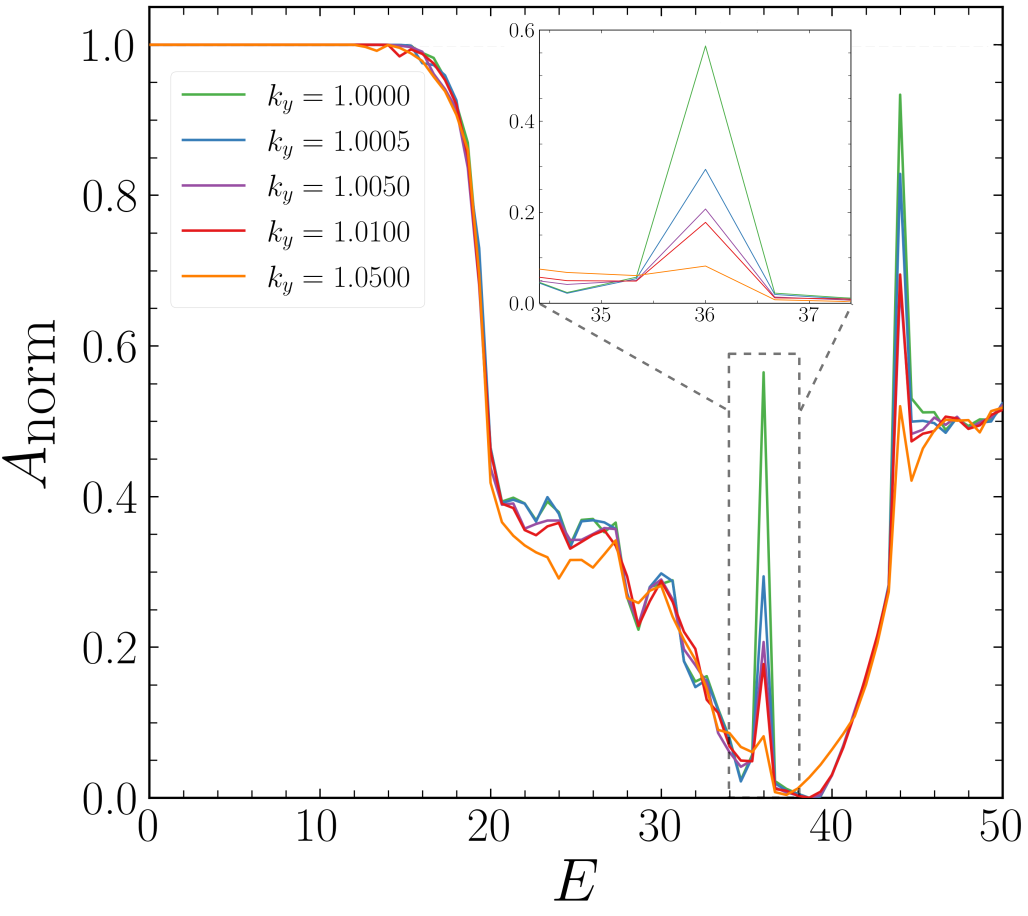}
    \caption{Regular area (normalized to total area $A_\textrm{total} = A_\textrm{regular} + 
             A_\textrm{chaos} = 1$) as a function of energy and fixed coupling ($\alpha=0.1$) for 
             increasing $k_y$ as asymmetry grows in the rectangular lattice.}
    \label{fig:chaotic-area-profile-var-ky}
\end{figure}

\begin{figure}
    \centering
    \includegraphics[width=0.5\textwidth]{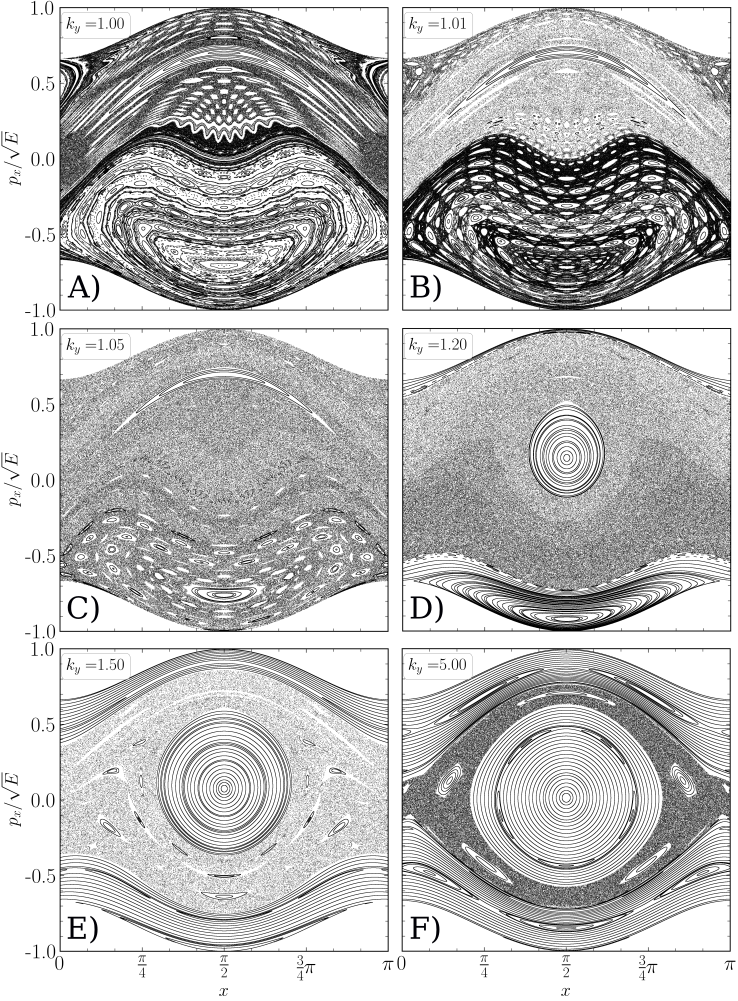}
    \caption{Poincar\'{e} section portraits of the island myriad for fixed energy ($E=36$) and coupling ($\alpha=0.1$) and 
             increasing $k_y$. \textbf{A)} $k_y = 1$ (square case); \textbf{B)} $k_y = 1.01$; \textbf{C)} $k_y = 1.05$;
             \textbf{D)} $k_y = 1.2$; \textbf{E)} $k_y = 1.5$; \textbf{F)} $k_y = 5$.}
    \label{fig:phase-space-rectangular-lattice}
\end{figure}

A verification of the myriad disappearance was carried out by measuring of the regular area profile 
for a fixed coupling value and varying energy, as the value of $k_y$ changes from the square case 
($k_x=k_y=1$) to asymmetric scenarios. 
Figure \ref{fig:chaotic-area-profile-var-ky} shows that, as the energy reaches the local maxima 
($E = V_\textrm{local} = 36$, for $\alpha=0.1$), the regular area presents a sudden peak, as expected 
for the square case $k_y=1$. 
As asymmetry grows with increasing $k_y$, this peak is quickly suppressed, with the myriad completely vanishing 
when $k_y \simeq 1.100$. This effect is also verified in phase-space portraits \textbf{A} to \textbf{C} in 
figure \ref{fig:phase-space-rectangular-lattice}, with the myriad being eroded by chaos. 
The same trend is seen for the myriad relative to the global maxima energy level ($E = V_\textrm{global} = 44$, 
for $\alpha=0.1$).

Furthermore, for even larger $k_y$ ($k_y > 1.2$), a stabilization is seen in phase-space for larger values of momentum, 
as shown in portraits \textbf{D} to \textbf{F}. Primarily for the bottommost region of the Poincar\'{e} section, 
for $p_x \approx -\sqrt{E}$, and later for the uppermost region $p_x \approx +\sqrt{E}$, islands and invariant curves 
appear and grow in area as the asymmetry between the $x$ and $y$ axes becomes more pronounced.
This indicates that the creation of a movement channel along the $y$ axis with a period different from the one along $x$ 
induces the stabilization of long flights, implying a pendulum-like dynamics along $x$ in the limit that its movement becomes 
uncoupled from the one in $y$.

\section{Island myriad -- hexagonal lattice}\label{sec:results-hexagonal}

\begin{figure}
    \centering
    \includegraphics[width=0.5\textwidth]{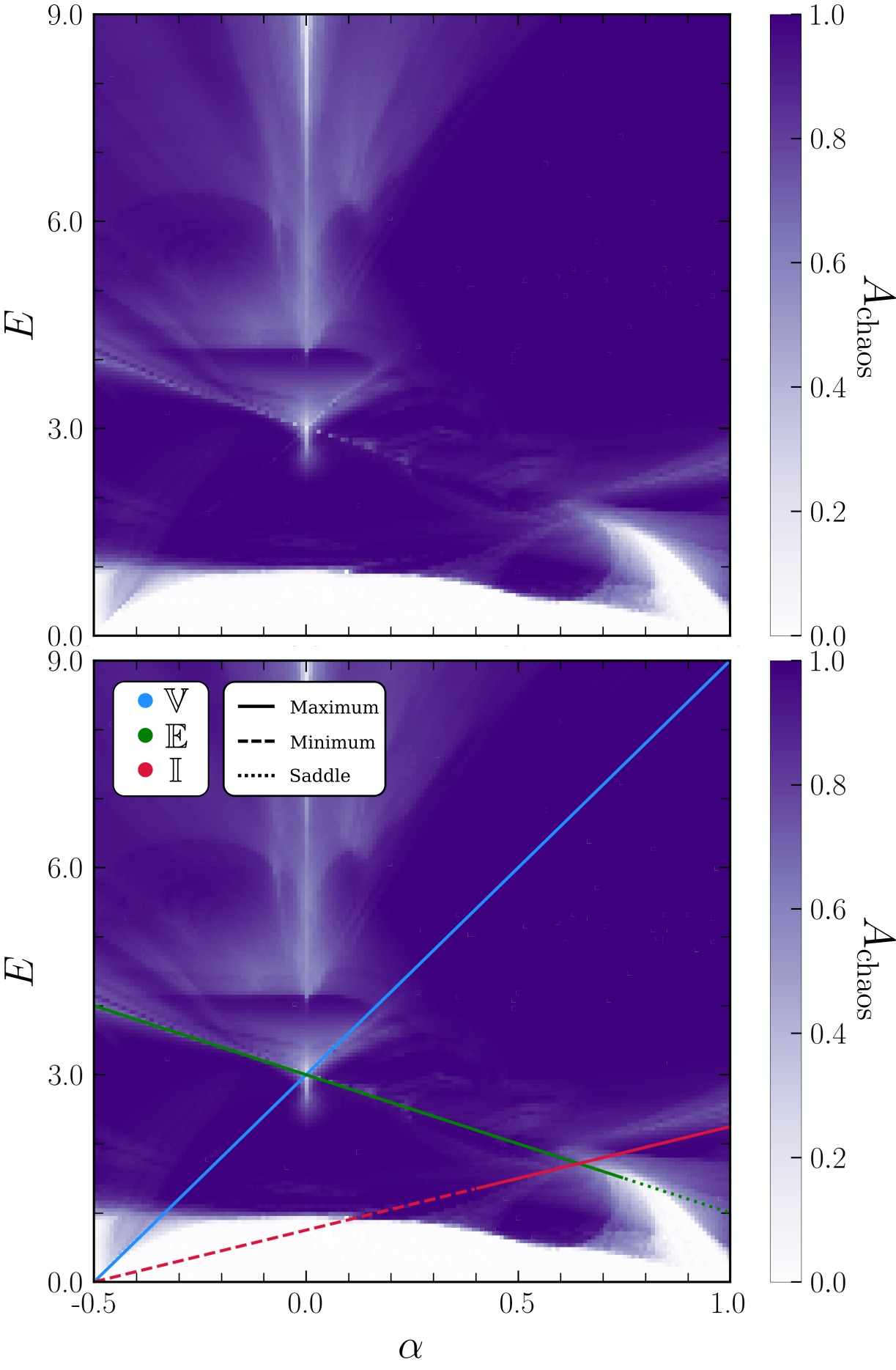}
    \caption{Color map of the chaotic area portion in parameter space for the hexagonal lattice. 
             Total chaos (regularity) is indicated by $A = 1$ ($A = 0$). In the 
             bottom frame, energy lines colors correspond to the equilibria 
             displayed following table \ref{tab:honeycomb-equilibrium-points} and the texture of the line 
             stands for the point stability. Grid size is $250 \times 250$.}
    \label{fig:chaotic-area-hexagonal}
\end{figure}

From the premise that the myriad relies on the potential function symmetries, we extend 
the investigation to a hexagonal system, as the next polygon with tiling property. 
Therewith, as done for the regular square system, the chaotic and regular area portions are shown 
in figure \ref{fig:chaotic-area-hexagonal}.

\begin{figure}
    \includegraphics[width=0.5\textwidth]{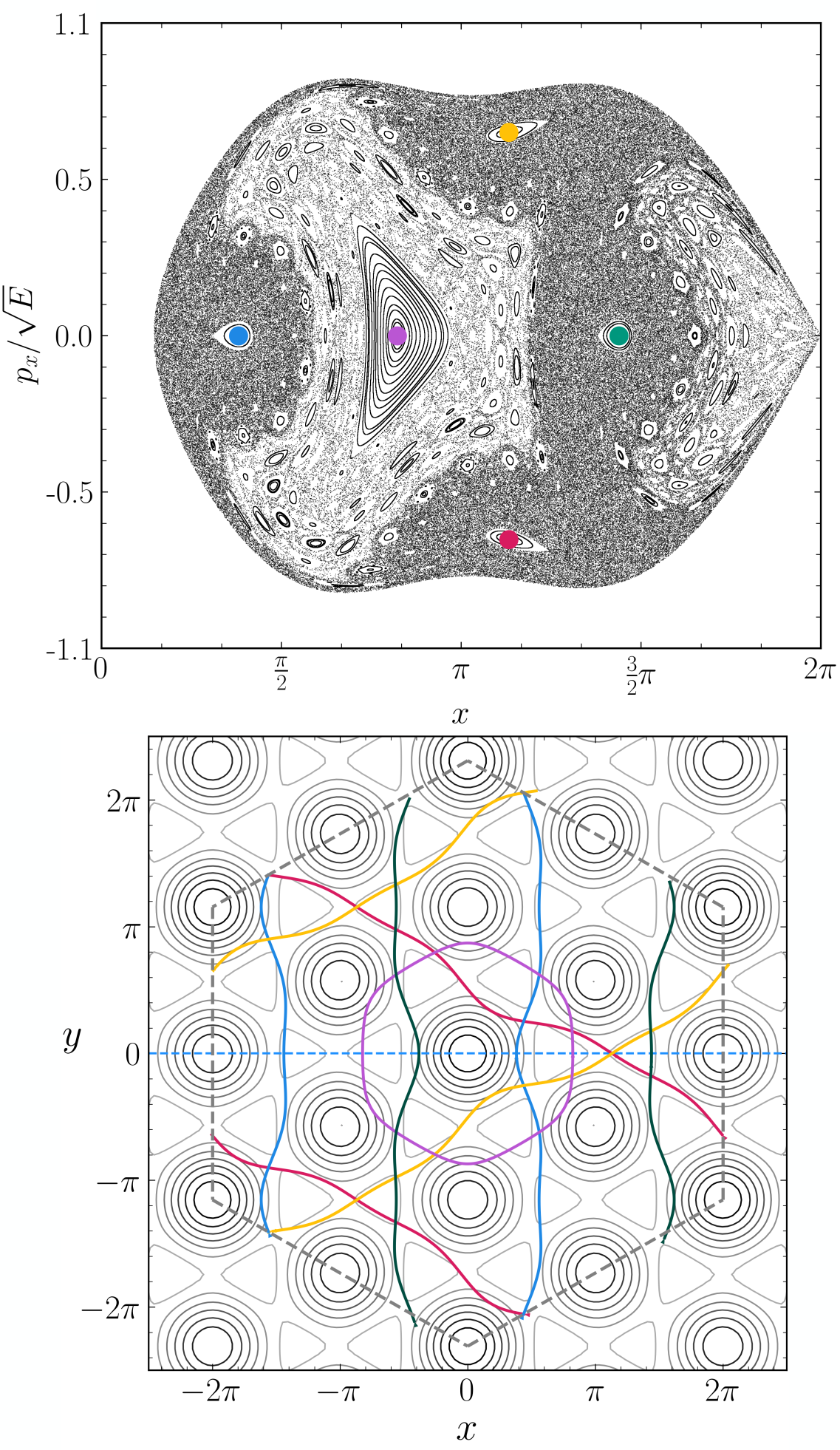}
    \caption{(Top) Island myriad in the hexagonal lattice as seen on the Poincar\'{e} section ($y=0; p_y > 0$) 
             for $\alpha = 0.04$ and $E = V(\mathbb{E}, \alpha) = 2.92$.
             (Bottom) Stable periodic orbits for the main stability islands shown in the hexagonal unit cell.}
    \label{fig:myriad-hexagonal-positive-alpha}
\end{figure}

\begin{figure}
    \includegraphics[width=0.5\textwidth]{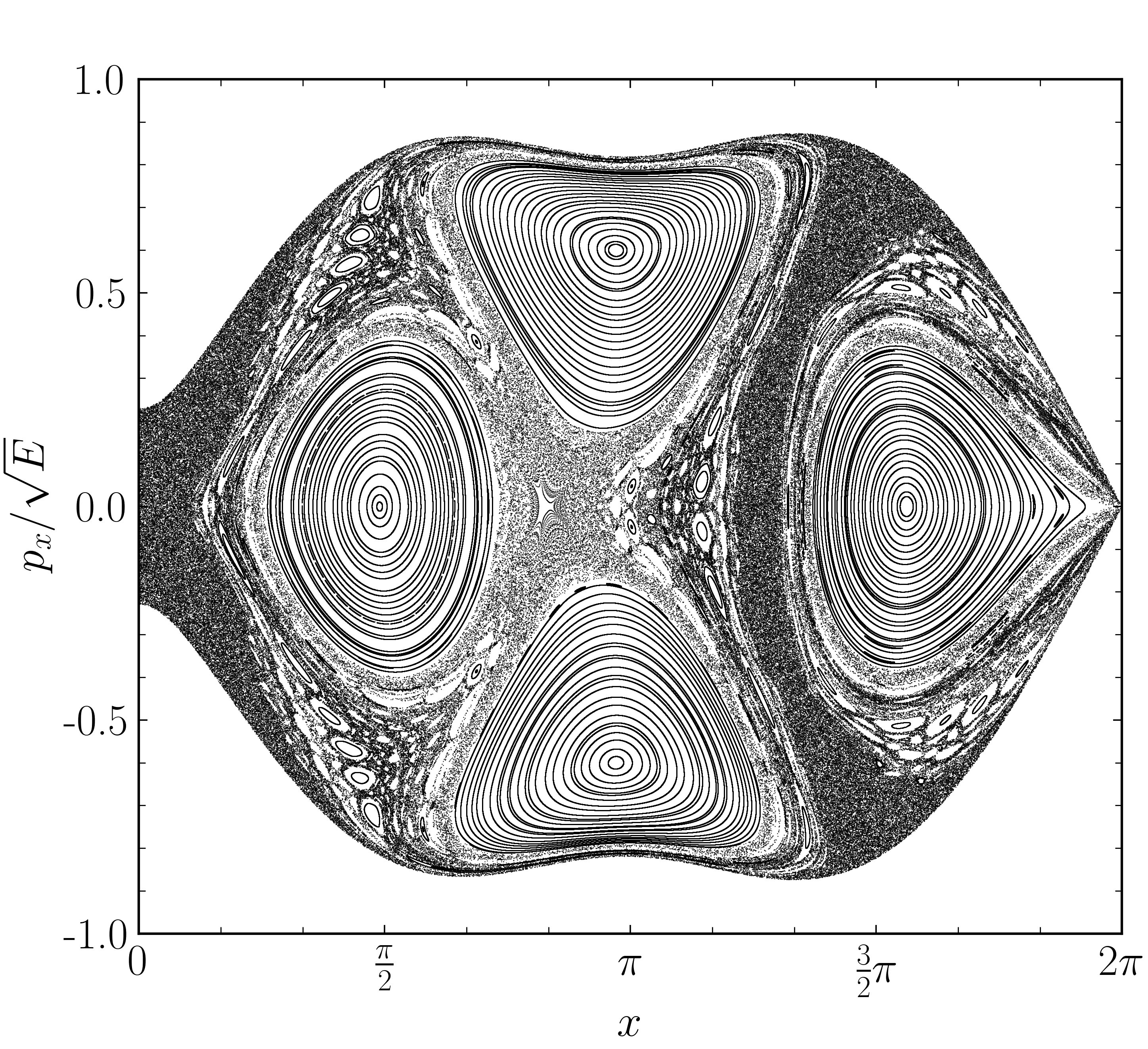}
    \caption{Island myriad in the hexagonal lattice as seen from the Poincar\'{e} section ($y=0; p_y > 0$) for 
             $\alpha = -0.02$ and $E = V(\mathbb{E}, \alpha) = 3.04$.}
    \label{fig:myriad-hexagonal-negative-alpha}
\end{figure}

As conjectured, the myriad is expected to emerge at energy levels 
of maxima of the potential surface.
However, in the hexagonal system, this correlation is not so prominent as in the square case. 
Indeed, the only region where it is clearly identified is near $\alpha \approx 0$, over the 
$V(\mathbb{E}, \alpha) = 3 - 2 \alpha$ line (in green in fig. \ref{fig:chaotic-area-hexagonal}). 
The myriad found is shown in phase-space in figures \ref{fig:myriad-hexagonal-positive-alpha} and 
\ref{fig:myriad-hexagonal-negative-alpha} for $\alpha \gtrsim 0$ and $\alpha \lesssim 0$, respectively. 
Despite the general similarity, for $\alpha \gtrsim 0$ the island chains surround only the center 
island, relative to a bounded periodic orbit (in purple in fig. \ref{fig:myriad-hexagonal-positive-alpha}), 
whereas for $\alpha \lesssim 0$ the island chains surround all 4 major islands.

At $\alpha = 0.0$, the parameter space reveals a vertical line with increased stable area seen for 
$E > 2.5$, as expected from a myriad structure. However, this increase was seen to be related to the 
stabilization of the 4 major islands shown in figures \ref{fig:myriad-hexagonal-positive-alpha} and 
\ref{fig:myriad-hexagonal-negative-alpha} and satellite islands, although not as a myriad. 
It becomes apparent then that at null coupling, where both $\mathbb{E}$ and $\mathbb{V}$ points are 
isoenergetic maxima, the inner triangulations within the unit cell, despite increasing symmetry, are 
not enough to form a myriad but instead changing $\alpha$ increases the stability area of the 4 main 
island orbits.

Despite the lack of visible fractality in the myriad as seen in the square system, the chaotic region 
in between chains present a strong stickiness behavior, acting as a permeable barrier for chaotic transport 
from the chaotic sea into the myriad core.
Indeed, the myriad in the hexagonal system seems to be affected by other instabilities in the potential 
surface caused by saddle and maxima points not listed here, preventing the existence or stabilization 
of periodic orbits to form the chains.

\begin{figure}
    \includegraphics[width=0.5\textwidth]{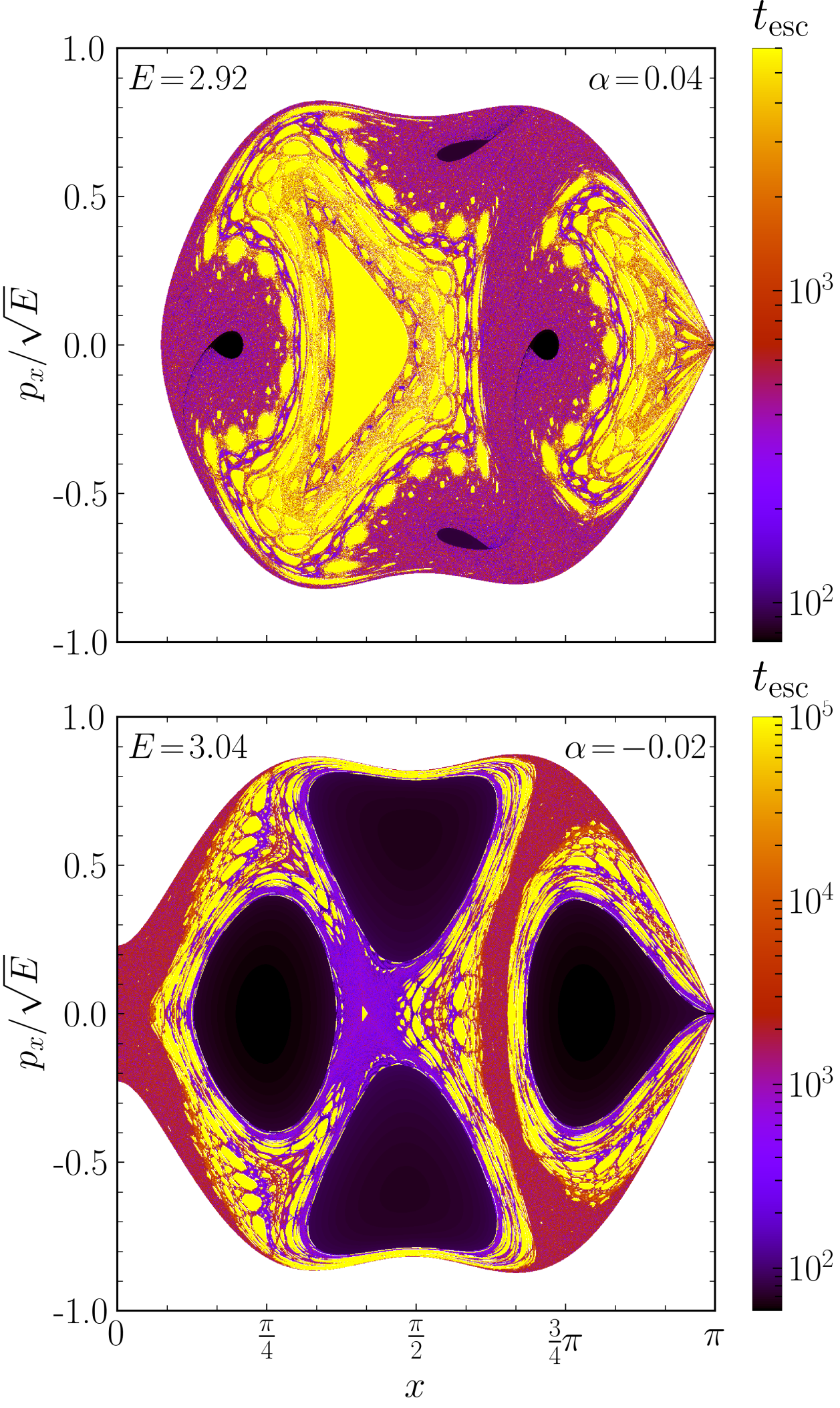}
    \caption{Escape time ($t_\textrm{esc}$) color map over the Poincar\'{e} section ($y = 0; p_y > 0$) 
             for the hexagonal lattice. 
             (Top) $E = 2.92$, $\alpha = 0.04$.
             (Bottom) $E = 3.04$, $\alpha = 2.92$.}
    \label{fig:hexagonal_escape}
\end{figure}

Besides the lack of pronounced fractality, the orbits comprising the myriad present less varied 
features regarding its periodic closure and isochronicity as compared to the ones seen in the square 
system. 
For example, figure \ref{fig:hexagonal_escape} shows the escape time pattern over the section 
from figure \ref{fig:myriad-hexagonal-positive-alpha}, revealing only trapped orbits (in yellow) through all myriad chains. 
Also, the stickiness in between chains become clearer once the chaotic region inside the myriad core has 
trapped orbits (up to time $t=6\times10^3$) despite being connected to the outer chaotic sea. 

\begin{figure}
    \includegraphics[width=0.5\textwidth]{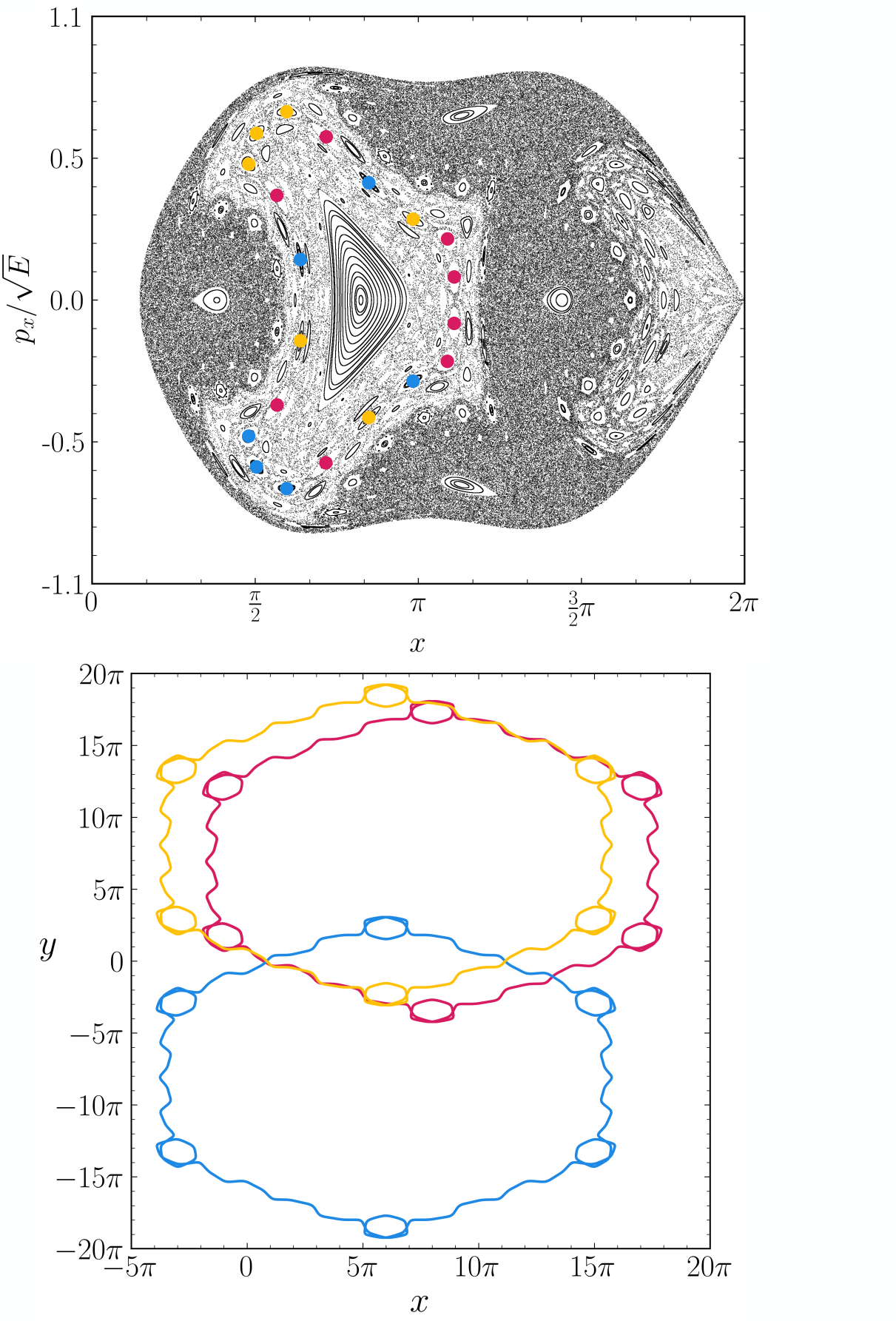}
    \caption{Poincar\'{e} section for $E = V(\mathbb{E},\alpha) = 2.92$ and $\alpha = 0.04$ with selected orbits from 
             a chain with total period $T = 20$. (Bottom) Isochronous periodic orbits without periodic boundary conditions: red ($T=8$); 
             blue ($T=6$); yellow ($T=6$). }
    \label{fig:hexagonal-isochronous-chain}
\end{figure}

\begin{figure}
    \includegraphics[width=0.5\textwidth]{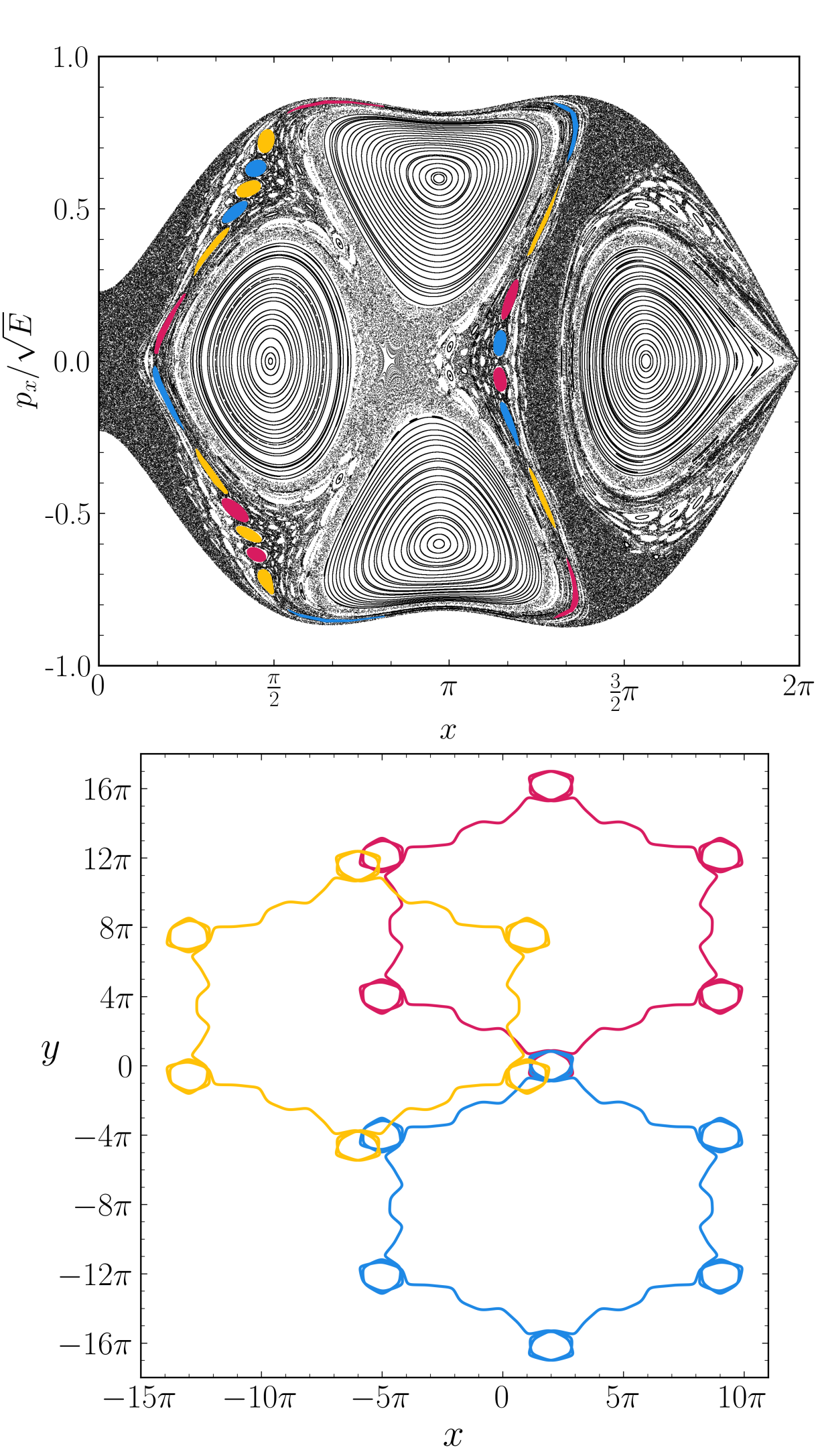}
    \caption{Poincar\'{e} section for $E = V(\mathbb{E},\alpha) = 3.04$ and $\alpha = -0.02$ with selected orbits from 
             a chain with total period $T = 22$. (Bottom) Isochronous periodic orbits without periodic boundary conditions: red ($T=7$); 
             blue ($T=7$); yellow ($T=8$). }
    \label{fig:hexagonal-isochronous-chain-2}
\end{figure}

Regarding isochronicity, once the hexagonal tiling has a three-fold rotation symmetry 
(from its three symmetry axes, $60^o$ apart from each other), the multiplicity of most chains 
is also three-folded, as exemplified in figures \ref{fig:hexagonal-isochronous-chain} and 
\ref{fig:hexagonal-isochronous-chain-2}. 
In this case, the orbits are invariant under rotations of $\pi / 3$, thus not altering their 
fixed point period when rotated. For this reason, isochronous orbits are simple translations 
from one another. 
In figure \ref{fig:hexagonal-isochronous-chain}, using the yellow orbit as a reference, 
the red orbit is translated in the $\hat{t}_{y \to r} = (\frac{\sqrt{3}}{2}, -\frac{1}{2})$ direction 
and the blue one along $\hat{t}_{y \to b} = (0, -1)$. Similarly in figure \ref{fig:hexagonal-isochronous-chain-2}, 
the yellow to red translation is along $\hat{t}_{y \to r} = (\frac{1}{2}, \frac{\sqrt{3}}{2})$ and 
the red to blue along $\hat{t}_{r \to b} = (0, -1)$.
As seen for the square system, higher multiplicity chains may occur, but none was found for this case. 

\section{Conclusions}\label{sec:conclusions}

Fundamentally, the island myriad is seen as the emergence of stability islands at energy levels of 
maxima in periodic potentials with tiling symmetry. 
Despite the instability of these equilibrium points, the periodic orbits deviated 
near these points are stable and appear as concentric layers of island chains in phase-space.
A thorough verification over parameter space reveals that this structure is exclusively found over a 
short energy interval near unstable points, being clearly visible along both local and global maxima 
for the square lattice, while being restricted to null coupling for the hexagonal case.

The myriad existence relies on the translational, rotational and mirror symmetries of the potential 
function. 
This dependence was directly verified when comparing the square lattice with its non-symmetric 
equivalent form as a rectangular lattice, with asymmetries of 5\% being enough to suppress the myriad 
($|k_y - k_x| = 0.05$). 
Moreover, further increasing the asymmetry between the $x$ and $y$ axes for $k_y \gg k_x$, the dynamics 
becomes uncoupled and phase-space is stabilized in a pendulum-like configuration, with chaos restricted 
to separatrix vicinity.

Particularly for the square lattice system, the myriad appears as a finite web-torus with notable fractality, 
where each chain has an even period split into 2 independent sets of isochronous orbits, 
although isolated cases with 3 sets were also found. 
Furthermore, the translational symmetry allows for different periodic closures, in 
the sense that some periodic orbits return to their initial position whereas others reach identical sites 
in translated unit cells by 
$(\Delta x, \Delta y) = (2m\pi, 2n\pi)$, for $m,n\in\mathbb{Z}$. 
The overall effect is the co-existence of trapped orbits (libration -- periodically closed) and long flights 
(rotation -- periodically open), therefore affecting global transport.
In addition to that, as the coupling parameter was increased, the myriad was seen to undergo 
separatrix reconnections for each of its layers, destroying them sequentially as they expand outwards the myriad core, 
suggesting a non-twist local dynamics over the Poincar\'{e} section.

One can add to these results our previous findings for the square system regarding the diffusive transport 
of particles \cite{Lazarotto}.
For the energy level of local maxima, the myriad emergence and sudden disappearance correlates to  
suppression in global diffusion, with long flights vanishing from the system dynamics. Also, periodic 
orbits approaching the maxima present a divergence in their period, as in the paradigmatic classic pendulum 
in its threshold between rotation and libration, therefore promoting a slowing down of the dynamics.

Despite being also confirmed in the hexagonal lattice, as expected from its similar tiling symmetries, 
the myriad was found in attenuated form. In this case, fractality is less pronounced due to extra  
saddle and maxima points in the potential surface acting as instability sources, 
thereby preventing the stabilization of orbits that would form the chains.
Also, only periodically closed orbits are found, thus with no simultaneous opposite transport regimes 
as in the square case.

Although isochronicity, separatrix reconnection and web-tori are already well-documented features 
of dynamical systems, lattice models present all of them simultaneously in a single structure. 
Moreover, the simplicity of the model allows for an intuitive understanding of these phenomena in 
phase-space and their correspondent dynamical behavior in position space. As seen from the periodic orbits that comprise 
the myriad chains, they are a direct consequence of the potential function symmetries, as opposed to more abstract 
models.

However, it is not yet intuitively clear why orbits in the myriad are found to be stable, where 
a more formal analytical description of the dynamics could better describe it. 
Zaslavsky \cite{Zaslavsky} presents a simple Hamiltonian for web-tori, although it does not contain 
the periodic closure and fractality properties seen here.

As indicated by the stable periodic orbits seen throughout this work, 
the myriad is formed as a consequence of scattered orbits approaching a set of unstable equilibria 
with equal energy. Baesens \textit{et al.} showed that, for chaotic scattering, 
these types of equilibria configuration give place to an abrupt bifurcation of hyperbolic 
orbits at energy levels close enough to the maxima in a smooth repulsive potential \cite{Baesens}.
From this premise, one could be inspired to justify the myriad as the consequence of 
similar bifurcations. 
But whereas Baesens \textit{et al.} consider a potential vanishing at infinity, 
in a lattice system, the tiling periodicity of $V(x,y)$ may provide stability to 
the orbits and therefore the appearance of the island chains along with their 
rotated and translated twin pairs.

In addition to that, a triangular lattice, the remaining polygonal periodic tiling shape, 
could present new features to the myriad phenomenon. 
Even though its construction cannot be achieved via the procedure used here, since only even-fold 
symmetry is allowed in the optical lattice setup, mathematically it could be promptly obtained.

\begin{acknowledgments}

M. Lazarotto would like to acknowledge Alexandre Poy\'{e} for fruitful discussions on the parallel 
computation of parameter spaces. 
We are indebted to anonymous reviewers for their constructive comments and suggestions. 
We acknowledge the financial support from the scientific agencies: 
S\~{a}o Paulo Research Foundation (FAPESP) under Grant No. 2018/03211-6;
Conselho Nacional de Desenvolvimento Cient\'{i}fico e Tecnol\'{o}gico (CNPq) under Grants No. 200898/2022-1 
and 304616/2021-4.
Coordena\c{c}\~{a}o de Aperfei\c{c}oamento de Pessoal de N\'{i}vel Superior 
(CAPES) and Comit\'{e} Fran\c{c}ais d'\'{E}valuation de
la Coop\'{e}ration Universitaire et Scientifique avec le
Br\'{e}sil (COFECUB) under Grant CAPES/COFECUB 8881.143103/2017-1.
Centre de Calcul Intensif d'Aix-Marseille is also acknowledged for granting 
access to its high-performance computing resources. 
\end{acknowledgments}

The data that support the findings of this study are available from the corresponding author upon reasonable request.

The authors declare to have no conflicts of interest to disclose.

\appendix

\section{Single coupling parameter restriction to the hexagonal lattice}\label{sec:append:single-coupling}

When assuming the single coupling condition for the hexagonal lattice, it is required to check whether 
it is feasible physically, as the couplings $\alpha_{nm}$ can be related to each other geometrically. 
Following Porter \textit{et al.} \cite{Porter1}, by assuming the first wave polarization versor 
$\hat{e}_1$ along the $\hat{z}$ direction, the remaining ones can be written in terms of spherical angles 
($\theta_j, \phi_j$) as
\nolinebreak
\begin{equation*}
\begin{cases}
    \hat{e}_1 = \hat{z} \\
    \hat{e}_2 = \cos(\phi_2)\sin(\theta_2) \;\hat{x} + \sin(\phi_2)\sin(\theta_2) \;\hat{y} + \cos(\theta_2) \;\hat{z} \\
    \hat{e}_3 = \cos(\phi_3)\sin(\theta_3) \;\hat{x} + \sin(\phi_3)\sin(\theta_3) \;\hat{y} + \cos(\theta_3) \;\hat{z},
\end{cases}
\end{equation*}
with $\phi_j \in [0, 2\pi)$ and $\theta_j \in [0, \pi]$. The couplings thus are
\begin{equation*}
\begin{cases}
    \alpha_{12} \!&= \hat{e}_1 \cdot \hat{e}_2 = \cos(\theta_2) \\
    \alpha_{13} \!&= \hat{e}_1 \cdot \hat{e}_3 = \cos(\theta_3) \\
    \alpha_{23} \!&= \hat{e}_2 \cdot \hat{e}_3 = \sin(\theta_2)\sin(\theta_3) \lt[ \cos(\phi_2)\cos(\phi_3) \rt. \\
                  &\hspace{1.5cm}                                             \lt.+ \sin(\phi_2)\sin(\phi_3)\rt]
                                                                            + \cos(\theta_2)\cos(\theta_3).
\end{cases}
\end{equation*}

When imposing the same value for all $\alpha_{nm}$, it must hold that $\theta_2 = \theta_3 = \theta$, 
implying the equality
\begin{equation*}
    \sin^2(\theta) \cos(\phi_2 - \phi_3) + \cos^2(\theta) = \alpha,
\end{equation*}
whence
\begin{equation*}
    \cos(\phi_2 - \phi_3) = \frac{\alpha}{1 + \alpha},
\end{equation*}
which will have real solutions $\phi_i$ only if $\lt|\frac{\alpha}{1 + \alpha}\rt| \leq 1$, thereby restraining 
$\alpha \in \lt[-\hlf, 1\rt]$. In short, it will only be possible to set $\alpha_{nm} = \alpha, \;\forall i, j$ 
by selecting $\theta_2 = \theta_3 = \theta$, such that $\cos(\theta) = \alpha$, and selecting values 
of $\phi_2, \phi_3$ such that $\cos(\phi_2 - \phi_3) = \lt(\frac{\alpha}{1 + \alpha}\rt)$, 
for $\alpha \in \lt[-\hlf, 1\rt]$.

\vspace{-0.6cm}
\section{Island myriad over global maxima}\label{sec:append:myriad-square-global-max}

\begin{figure}
    \centering
    \includegraphics[width=0.5\textwidth]{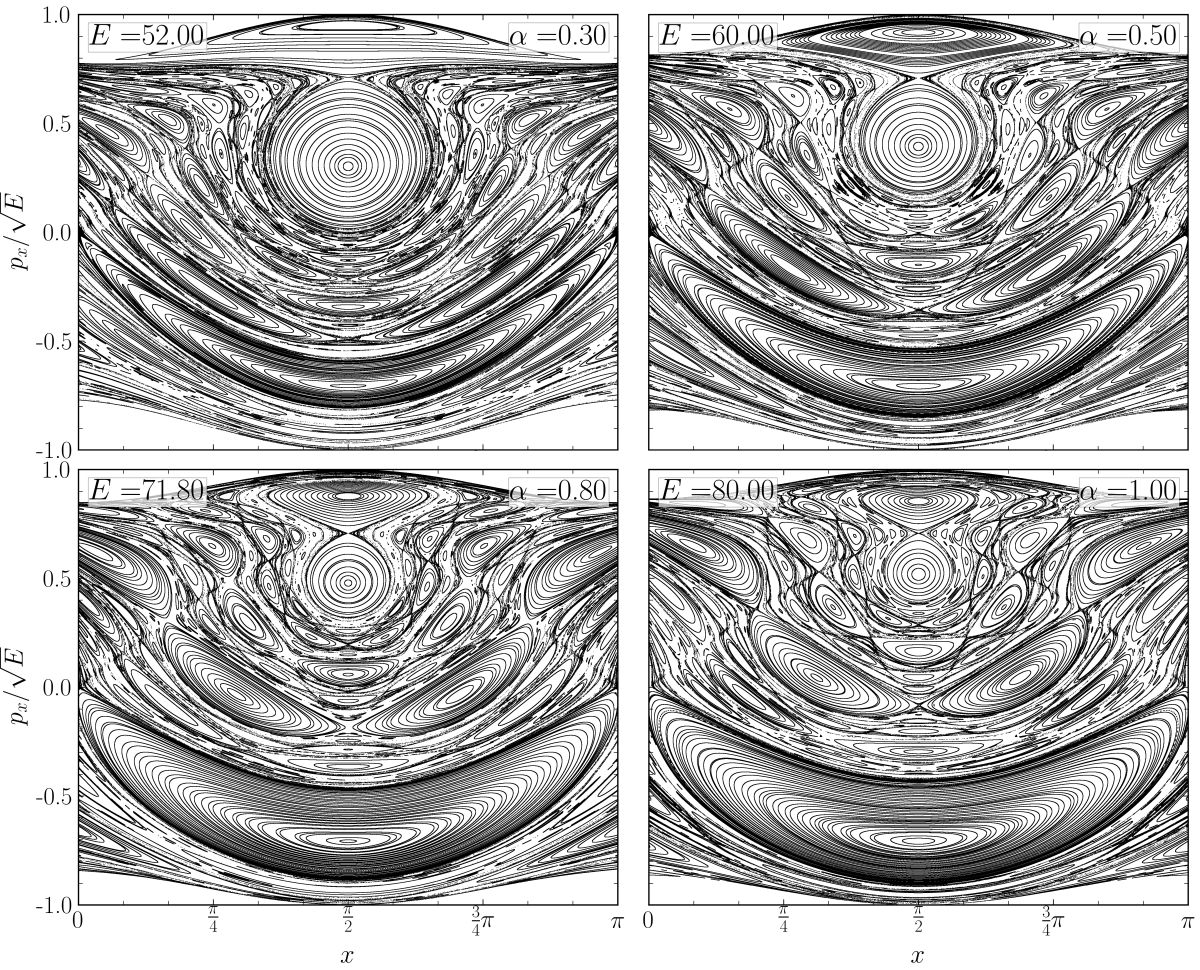}
    \caption{Section $\Sigma$ calculated for energy values at the global maxima energy line, 
             showing the island myriad for different couplings $\alpha$ for the square lattice.}
    \label{fig:myriad-global-maxima}
\end{figure}

Figure \ref{fig:myriad-global-maxima} shows different portraits of the island myriad for the square lattice 
at energy values over the potential global maximum $V_\textrm{g-max} = 2 U (1 + \alpha)$. The emergent structure is 
qualitatively similar for any $\alpha$ considered, with the size of resonant islands increasing with the coupling.

\section{Triple folded isochronicity}\label{sec:append:triple-isochronicity}

\begin{figure}[H]
  \centering
  \includegraphics[width=0.5\textwidth]{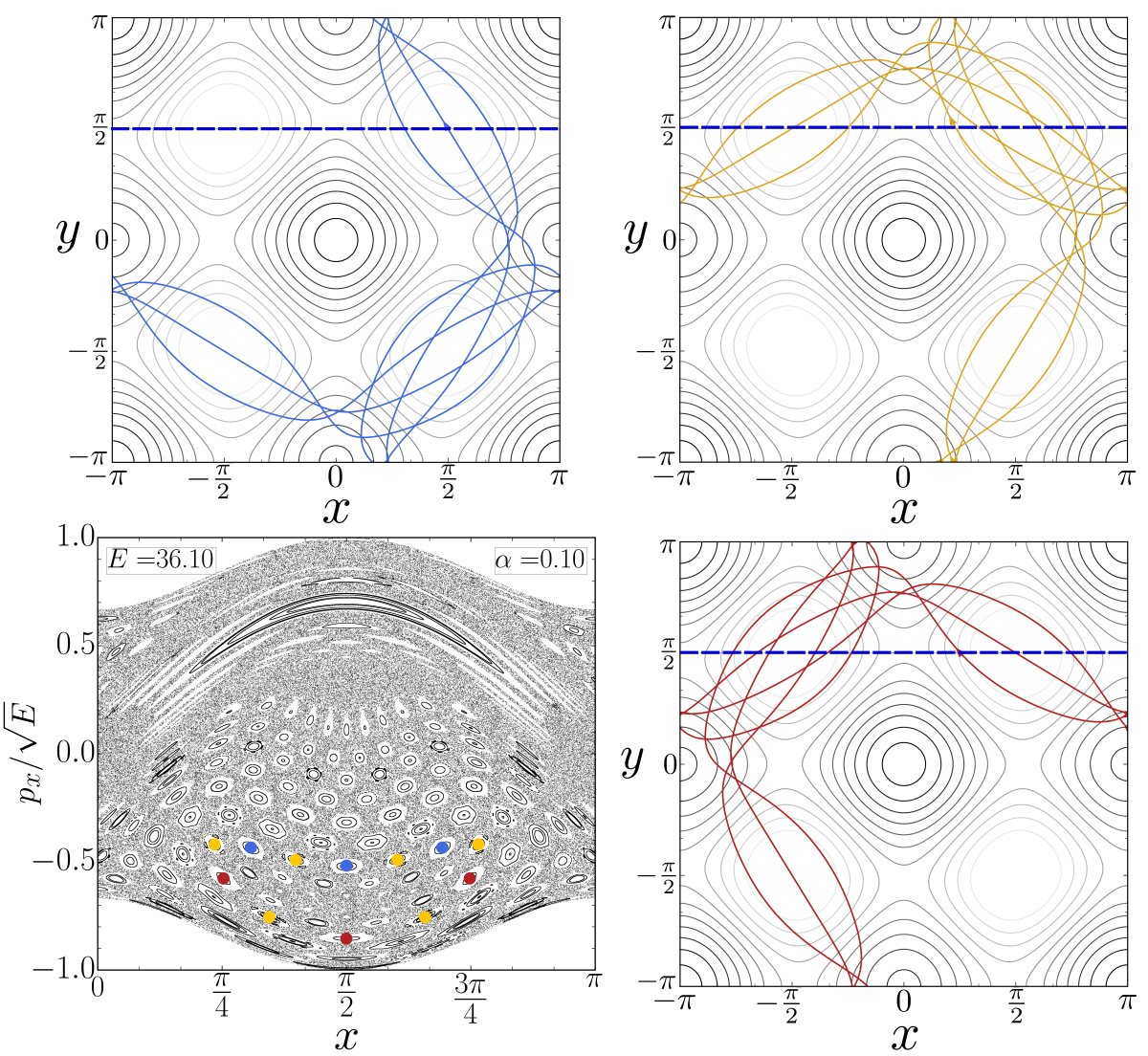}
  \caption{Single myriad chain formed by three isochronous orbits for the square lattice. The colored 
           dots indicate the fixed points of the trajectories relative to the section $\Sigma$ (blue 
           dotted line in trajectory frames).}
  \label{fig:triple-isochronicity}
\end{figure}

Figure \ref{fig:triple-isochronicity} shows a scenario for the square lattice where a single isochronous 
chain, with 12 islands, is formed not by two sets of period 6 chains, but instead by two sets of 
period 3 (shown in red and blue) and one of period 6 (shown in yellow). The orbits themselves show 
that they are indeed the same curve rotated and mirrored in three different ways, revealing that 
whenever an orbit's translation or rotation intersects the Poincar\'{e} section with the same discrete period, higher 
multiplicities may appear. 
However, for the square lattice, no more than 4 isochronous sets can be expected to appear, since its 
symmetries are limited by rotations of a quarter of cycle ($\frac{\pi}{2}$).

\pagebreak
\bibliographystyle{unsrt}
\bibliography{ref.bib}

\end{document}